\documentclass[%
 aip,
 amsmath,amssymb,jcp,
preprint,%
]{revtex4-1}

\usepackage{graphicx}
\usepackage{dcolumn}
\usepackage{bm}
\usepackage{color}
\usepackage[utf8]{inputenc}
\usepackage[T1]{fontenc}
\usepackage{mathptmx}

\begin{document}
\sloppy
\title{Efficient equilibration of confined and free-standing films of highly entangled polymer melts}

\author{Hsiao-Ping Hsu}
\email[]{hsu@mpip-mainz.mpg.de}
\affiliation{Max-Planck-Institut f\"ur Polymerforschung, Ackermannweg 10, 55128, Mainz, Germany}
\author{Kurt Kremer}
\email[]{kremer@mpip-mainz.mpg.de}
\affiliation{Max-Planck-Institut f\"ur Polymerforschung, Ackermannweg 10, 55128, Mainz, Germany}


\begin{abstract}
Equilibration of polymer melts containing highly entangled long polymer chains in confinement or with free surfaces is a challenge for computer simulations. We approach this problem by first studying polymer melts based on the soft-sphere coarse-grained model confined between two walls with periodic boundary conditions in two directions parallel to the walls. Then we apply backmapping to reinsert the microscopic details of the underlying bead-spring model. Tuning the strength of the wall potential, the monomer density of confined polymer melts in equilibrium is kept at the bulk density even near the walls. In a weak confining regime, we observe the same conformational properties of chains as in the bulk melt showing that our confined polymer melts have reached their equilibrated state. Our methodology provides an efficient way of equilibrating large polymer films with different thicknesses and is not confined to a specific underlying microscopic model. Switching off the wall potential in the direction 
perpendicular to the walls, enables to study free-standing highly entangled polymer films or polymer films with one supporting substrate.
\end{abstract}


\maketitle

\section{Introduction}

Polymer confinement plays an important role for many aspects of adhesion, wetting, lubrication, and friction of complex fluids from both theoretical and technological points of view. For more than two decades, both theoretical and experimental works~\cite{Thompson1992,Eisenriegler1993,Aoyagi2001,Harmandaris2005,Batistakis2012,FKremer2014,Pressly2019} have shown that the dynamic and structural properties of polymers subject to confinement may deviate from that in the bulk as the interaction between polymers and confining surfaces becomes non-negligible. For example, the conformations of polymer chains in a melt near the wall in the direction perpendicular to the wall shrink remarkably compared to bulk chains while they only extend slightly parallel to the wall~\cite{Aoyagi2001,Pakula1991,Cavallo2005}. Long chain mobility in confined melts is also affected by entanglement effects while in turn the conformational deviations from bulk chains have influence on the distribution of entanglements~\cite{Shin2007,Sussman2014,Russell2017,Lee2017,Garcia2018}. Furthermore many studies have also focused on the dependency between the glass transition temperature $T_g$ and the nature of the confinement effects~\cite{Alcoutlabi2005,Ediger2014,Vogt2018}. Therefore, it is important to understand the mechanical properties of confined polymer melts and how confinement impacts both viscous and elastic properties of amorphous polymer films with different surface substrates or even free surfaces.

Computer simulations provide a powerful method to mimic the behavior of polymers under well-defined external conditions covering the range from atomic to coarse-grained (CG) scales~\cite{Binder2005,Barrat2010}. However, the cost of computing time rises dramatically as the size and complexity of systems increase. Therefore, applying appropriate coarse-grained models that keep the global thermodynamic properties and the local mechanical and chemical properties is still an important subject~\cite{Murat1998,Kremer1990,Kremer1992,Plathe2002,Harmandaris2006,Gujrati2010,Vettorel2010,Zhang2013,Karatrantos2019}. One of the successful monomer-based models, namely the bead-spring (BS) model~\cite{Kremer1990,Kremer1992} together with an additional bond-bending potential~\cite{Faller1999,Faller2000,Faller2001,Everaers2004}, has been successfully employed to provide a better understanding of generic scaling properties of polymer melts in bulk. For such a model static and dynamic properties of highly entangled polymer melts in bulk have been extensively studied in our previous work~\cite{Hsu2016,Hsu2017}. We have verified the crossover scaling behavior of the mean square displacement of monomers between several characteristic time scales as predicted by the Rouse model~\cite{Rouse1953} and reptation theory~\cite{deGennes1979,Doi1980,Doi1983,Doi1986} over several orders of magnitude in the time. For weakly semiflexible polymer chains of sizes of up to $N=2000$ monomers we have also confirmed that they behave as ideal chains to a very good approximation. For these chains the entanglement length $N_e=28$ in the melt, estimated through the primitive path analysis and confirmed by the plateau modulus~\cite{Everaers2004,Moreira2015,Hsu2016}, resulting in polymer chains of $N=2000 \approx 72 N_e$. Thus, we here focus on this model and a related variant for the study of the equilibration of polymer melts confined between two repulsive walls as supporting films, and free-standing films after walls are removed. Each film contains $n_c=1000$ chains of $N=2000$ monomers at the bulk melt density {$\rho_0=0.85\sigma^{-3}$}. Directly equilibrating such large and highly entangled chains in bulk or confinement is not feasible within reasonably accessible computing time.  

A novel and very efficient methodology has recently been developed~\cite{Zhang2013,Zhang2014} for equilibrating large and highly entangled polymer melts in bulk described by the bead-spring model~\cite{Kremer1990,Kremer1992}. Through a hierarchical backmapping of CG chains described by the soft-sphere CG model~\cite{Vettorel2010,Zhang2013} from low resolution to high resolution and a reinserting of microscopic details of bead-spring chains, finally, highly entangled polymer melts in bulk are equilibrated by molecular dynamics (MD) simulations using the package ESPResSO++~\cite{Espressopp,Espressopp20}. To first order, the required computing time depends only on the overall system size and becomes independent of chain length. Similar methodologies have also been used to equilibrate high-molecular-weight polymer blends~\cite{Ohkuma2018} and polystyrene melts~\cite{Zhang2018}. In this paper, we extend the application of the soft-sphere approach to confined polymer melts and subsequently free-standing films. As polymer chains are described at a lower resolution, the number of degrees of freedom becomes smaller. Here we adapt this hierarchical approach to equilibrate polymer melts confined between two walls in detail. Moreover, we apply our newly developed, related model~\cite{Hsu2019,Hsu2019e} to prepare polymer films with one or two free surfaces. Differently from Refs.~\onlinecite{Vettorel2010,Zhang2014}, we take the bending elasticity of bead-spring chains in a bulk melt into account for the parameterization of the soft-sphere CG model. Namely, the underlying microscopic bead-spring chains are weakly semiflexible (bending constant $k_\theta=1.5\epsilon$) instead of fully flexible ($k_\theta=0.0\epsilon$). 

The outline of this paper is as follows: In Sec.~II, we introduce the main features of the microscopic bead-spring model and soft-sphere coarse-grained model used for studying the confined polymer melts. The application of the soft-sphere CG model for confined coarse-graining melts, and conformational properties of fully equilibrated confined CG melts are addressed in Sec.~III. In Sec.~IV, we reinsert the microscopic details of confined CG melts, and discuss the equilibration procedures. In Sec.~V, we show how to prepare films with one or two free surfaces by switching to another variant of bead-spring model. Finally, our conclusion is given in Sec.~VI.

\section{Models}
\subsection{Generic microscopic bead-spring models}
\label{BSM}

In the microscopic bead-spring model~\cite{Kremer1990,Kremer1992}, all monomers at a distance $r$ interact via a shifted Lennard-Jones (LJ) potential $U_{\rm LJ}(r)$,
\begin{eqnarray}
U_{\rm LJ}(r)=\left\{\begin{array}{ll}
V_{\rm LJ}(r)-V_{\rm LJ}(r=r_{\rm cut})    & \;, \, r \le r_{\rm cut} \\
0 &\;, \, r>r_{\rm cut} 
\end{array} \right.  \, ,
\label{eq-ULJ}
\end{eqnarray}
with
\begin{equation}
V_{\rm LJ}(r)=4\epsilon\left[\left(\frac{\sigma}{r}\right)^{12}-\left(\frac{\sigma}{r}\right)^{6}
\right]
\end{equation}
where $\epsilon$ is the energy strength of the pairwise interaction, $r_{\rm cut}$ is the cutoff in the minimum of the potential such that force and potential are zero at $r_{\rm cut}=2^{1/6}\sigma$. These LJ units also provide a natural time definition via $\tau= \sigma \sqrt{m/\epsilon}$, $m=1$ being the mass of the monomers. The temperature is set to $T = 1.0\epsilon/k_B$, $k_B$ being the Boltzmann factor, which is set to one. Along the backbone of the chains a finitely extensible nonlinear elastic (FENE) binding potential $U_{\rm FENE}(r)$ is added,
\begin{eqnarray}
U_{\rm FENE}(r) = \left\{\begin{array}{ll}
-\frac{k}{2}R_0^2 \ln \left[1-\left(\frac{r}{R_0}\right)^2 \right] & \;,\, r <  R_0 \\
\infty & \;, \, r \ge R_0 
\end{array} \right . \, ,
\label{eq-UFENE}
\end{eqnarray}
where $k=30 \epsilon/\sigma^2$ is the force constant and $R_0=1.5 \sigma$ is the maximum value of bond length. For controlling the bending elasticity, i.e., chain stiffness, the standard bond-bending potential~\cite{Faller1999,Faller2000,Faller2001,Everaers2004} is given by
\begin{equation}
U_{\rm BEND}^{\rm (old)}(\theta)=k_\theta (1-\cos \theta) \,, \qquad 0<\theta <\pi
\label{eq-UBENDold}
\end{equation}
with the bond angle $\theta$ defined by
$
\theta=\cos^{-1}\left(  \frac{\vec{b}_j \cdot \vec{b}_{j+1}}
{\mid \vec{b}_j \mid \mid \vec{b}_{j+1} \mid} \right)
$
where $\vec{b}_j=\vec{r}_{j+1}-\vec{r}_{j}$ is the bond vector between the $(j+1)$th monomer and the $j$th monomer along the identical chain. The bending factor $k_\theta$ is set to $1.5\epsilon$ and the melt density is set to the widely used value of $0.85 \sigma^{-3}$ throughout the whole paper. 

{For studying polymer melts under confinement, w}e first consider the simpler example of polymer melts that are confined between two planar, structureless repulsive walls. The walls placed at $z=0$ and $z=L_z$ are described by the 10-4 Lennard-Jones planar wall potential ~\cite{Grest1996,Aoyagi2001},
\begin{eqnarray}
    U_{\rm wall}(z)=\left\{\begin{array}{ll}
V_{\rm wall}(z)-V_{\rm wall}(z=\sigma) \quad &,\quad  z \leq \sigma \\
V_{\rm wall}(L_z-z)-V_{\rm wall}(L_z-z=\sigma) \quad &,\quad  L_z-z \leq \sigma \\ 
0  &,\quad {\rm otherwise}
\end{array} \right .
\label{eq-Uwall}
\end{eqnarray}
with
\begin{equation}
  V_{\rm Wall}= 4 \pi \varepsilon_w \left [\frac{1}{5}\left(\frac{\sigma}{z}\right)^{10}-\frac{1}{2}
\left(\frac{\sigma}{z}\right)^4 \right] \,.
\label{eq-Vwall}
\end{equation}
Here $\varepsilon_w$ is the interaction strength between {monomers} and the walls, $z$ and $L_z-z$ are the vertical distances of a monomer from the two walls, respectively.

For preparing polymer films with one or two free surfaces, we have to stabilize the system at zero pressure, particularly, in the direction perpendicular to the walls. Only then we can switch off the wall potential and prevent system instability. Therefore, a short-range attractive potential to reduce the pressure to zero is added~\cite{Hsu2019,Hsu2019e} with an additional shift term,
\begin{eqnarray}
U_{\rm ATT}(r)=\left\{\begin{array}{ll}
\alpha \left[ \cos(\pi \left(\frac{r}{r_{\rm cut}} \right)^2) +1 \right] 
,& {r_{\rm cut}} \leq r < r^a_{c} \\
& \\
0  ,& {\rm otherwise}
\end{array} \right .  \,,
\label{eq-Uatt}
\end{eqnarray}
such that $U_{\rm ATT}(r)=U_{\rm LJ}(r)$ at $r=r_{\rm cut}$. Here $\alpha=0.5145\epsilon$ is the strength parameter, $r^a_{c}=\sqrt{2}r_{\rm cut} \approx 1.5874\sigma$ is the upper cut-off such that it has zero force at $r=r_{\rm cut}$ and $r=r^a_{c}$. Note that this additional potential does not alter the characteristic conformations at $T= 1 \epsilon / k_B$, so that we can switch between these different models as needed~{\cite{Kremer1990,Kremer1992,Auhl2003,Hsu2019,Hsu2019e}}.

{Note that using $U_{\rm BEND}^{\rm (old)}(\theta)$, bead-spring chains tend to stretch out as the temperature decreases. To avoid such an artificial chain stretching, Eq.~(\ref{eq-UBENDold}) can be replaced by~\cite{Hsu2019,Hsu2019e}
\begin{equation}
U_{\rm BEND}(\theta) = -a_\theta \sin^2 (b_\theta \theta) 
\,, \qquad 0 < \theta <\theta_c=\pi/b_\theta \,
\label{eq-UBEND}
\end{equation}
if one is interested in studying polymer melts under cooling. The fitting parameters $a_\theta=4.5\epsilon$ and $b_\theta=1.5$ are determined so that the local conformations of chains remain essentially unchanged compared to those with $k_\theta=1.5\epsilon$ using Eq.~(\ref{eq-UBENDold}) at temperature $T=1.0\epsilon/k_B$.}

\subsection{Soft-sphere coarse-grained model}

\begin{figure*}[t!]
\begin{center}
\includegraphics[width=0.70\textwidth,angle=0]{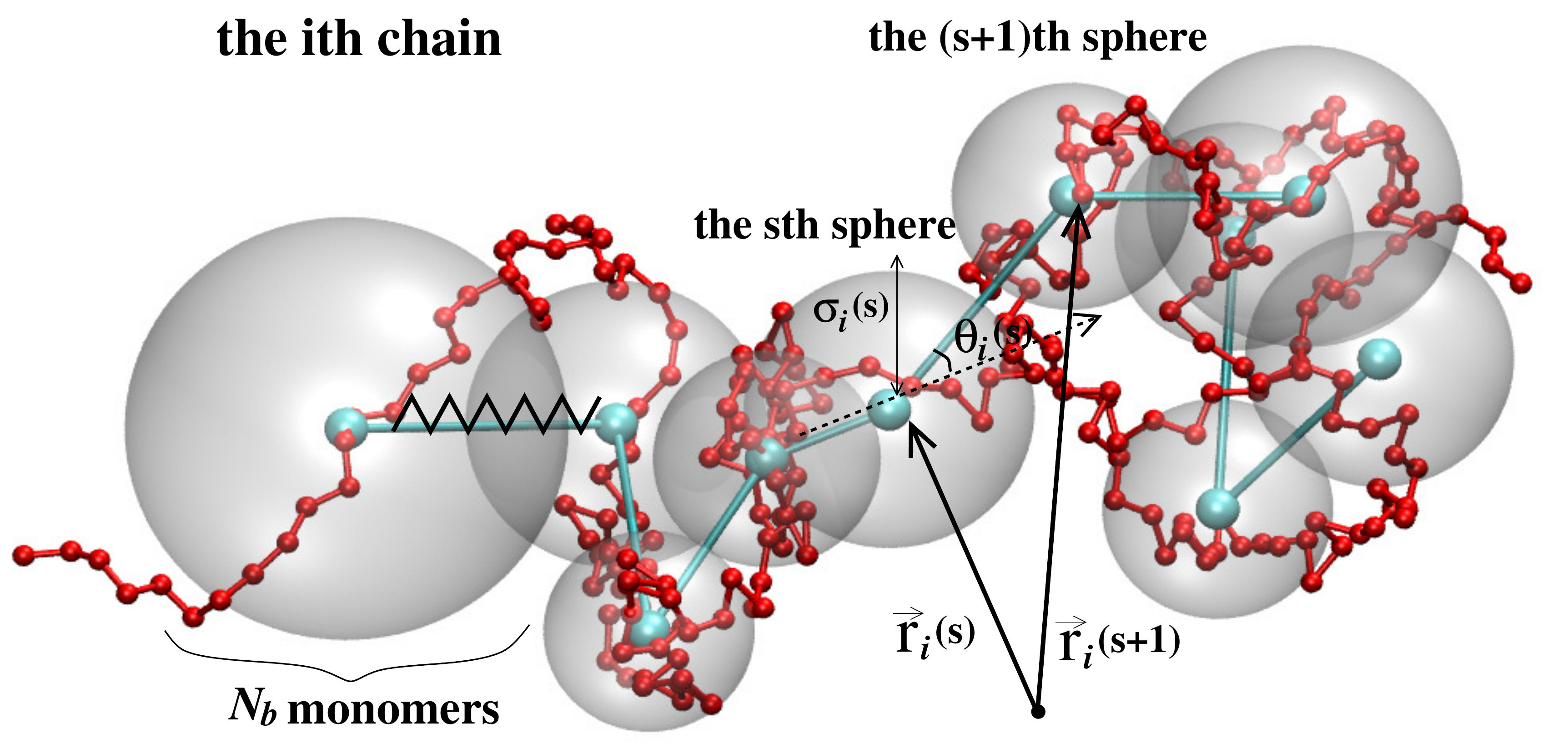}
\caption{Snapshot of the configuration of a bead-spring chain of $N=250$ monomers represented
by a coarse-grained chain of $N_{\rm CG}=10$ fluctuating soft spheres. Each sphere
corresponds to $N_b=25$ monomers, cf. text.}
\label{fig-bsmsph}
\end{center}
\end{figure*}

In the soft-sphere approach~\cite{Vettorel2010}, each bead-spring polymer chain in a melt is represented by a chain of $N_{\rm CG}=N/N_b$ fluctuating soft spheres in a CG view as shown in Fig.~\ref{fig-bsmsph}. The coordinate of the center and the radius of the $s$th sphere in the $i$th chain, $\vec{r}_i(s)$ and $\sigma_i(s)$, are described by
\begin{equation}
\vec{r}_i(s)=\vec{R}_{{\rm CM},i}(s)=\frac{1}{N_b}\sum_{k=(s-1)N_b+1}^{sN_b}\vec{r}_{i,k} 
\label{eq-mapping-cm}
\end{equation}
and
\begin{equation}
\sigma_i(s)=R_{g,i}(s)=\left(\frac{1}{N_b}\sum_{k=(s-1)N_b+1}^{sN_b}
\mid \vec{r}_{i,k}-\vec{r}_i(s) \mid^2 \right)^{1/2} \,,
\label{eq-mapping-rg}
\end{equation}
respectively. Here $\vec{r}_{i,k}$ is the coordinate of the $k$th monomer in chain $i$, $\vec{R}_{{\rm CM},i}(s)$ and $R_{g,i}(s)$ are the center of mass (CM) and the radius of gyration of $N_b$ monomers in its subchains, respectively. {Since we extend the application of the soft-sphere CG model to semiflexible chains and follow the slightly different strategies of Refs.~\onlinecite{Zhang2013,Zhang2014}, we list the details of soft-sphere potentials as follows.} The spheres are connected by a harmonic bond potential
\begin{equation}
   U_{\rm bond}^{\rm (CG)}(d_i(s)=\mid \vec{d}_i(s) \mid) =\frac{3}{2b^2_{\rm CG}}{d_i(s)^2}
\label{eq-Ubond-CG}
\end{equation}
and an angular potential
\begin{equation}
   U_{\rm ang}^{\rm (CG)}(\theta_i(s))=\frac{1}{2}k_{\rm bend}(1-\cos \theta_i(s))
\quad \textrm{with} \quad 
\cos \theta_i(s)=\frac{\vec{d}_i(s) \cdot \vec{d}_i(s+1)}
{\mid \vec{d}_i(s)\mid \mid \vec{d}_i(s+1) \mid}
\label{eq-Uang-CG}
\end{equation}
where $\vec{d}_i(s)=\vec{r}_i(s+1)-\vec{r}_i(s)$ is the bond vector, and $b_{\rm CG}$ denotes the effective bond length. The radius of the $s$th sphere in chain $i$ and its fluctuation are controlled by the following two potentials~\cite{Vettorel2010,Zhang2013},
\begin{equation}
   U_{\rm sphere}^{\rm (CG)}(\sigma_i(s))=a_1\frac{N_b^3}{\sigma^{6}_i(s)}+a_2\frac{\sigma_i^2(s)}{N_b}
\end{equation}
and
\begin{equation}
  U_{\rm self}^{\rm (CG)}(\sigma_i(s))=a_3\sigma_i^{-3} (s)\,,
\label{eq-Uself-CG}
\end{equation}
respectively. The non-bonded interactions between any two different spheres due to excluded volume interactions according to Flory's theory~\cite{Flory1949,deGennes1979,Murat1998} is taken care of by the potential
\begin{eqnarray}
&&  U_{\rm{nb}}^{\rm (CG)}(\vec{r}_i(s),\sigma_i(s);\vec{r}_{i'}(s'),\sigma_{i'}(s')) \nonumber \\
&&=\varepsilon_1 \left(\frac{2\pi(\sigma_i^2(s)+\sigma_{i'}^2(s'))}{3}\right)^{-3/2}
\exp\left[-\frac{3(\vec{r}_i(s)-\vec{r}_{i'}(s'))^2}{2(\sigma_i^2(s)+\sigma_{i'}^2(s'))}
\right] \, \textrm{for} \, i\neq i' \,\textrm{or} \, s \neq s'\,.
\label{eq-Unb-CG}
\end{eqnarray}
Here the parameters $k_{\rm bend}$, $a_1$, $a_2$, $a_3$, $b_{\rm CG}$, and $\varepsilon_1$ depending on $N_b$ are determined by a numerical approach {via curve fitting.}

Similar as shown in Eq.~(\ref{eq-Uwall}), the two soft repulsive planar walls located at $z=0$ and $z=L_z$ depending on the radius of each sphere is given by
\begin{eqnarray}
    U_{\rm wall}^{\rm (CG)}(z)=\left\{\begin{array}{ll}
V_{\rm wall}^{\rm (CG)}(z)-V_{\rm wall}^{\rm (CG)}(z=\sigma_i(s)) \quad &,\quad  z \leq \sigma_i(s) \\
V_{\rm wall}^{\rm (CG)}(L_z-z)-V_{\rm wall}^{\rm (CG)}(L_z-z=\sigma_i(s)) \quad &,\quad  L_z-z \leq \sigma_i(s) \\ 
0  &,\quad {\rm otherwise}
\end{array} \right .
\label{eq-Uwall-CG}
\end{eqnarray}
with
\begin{equation}
  V_{\rm Wall}^{\rm (CG)}= 4 \pi \varepsilon_w^{\rm (CG)} 
\left [\frac{1}{5}\left(\frac{\sigma_i(s)}{z}\right)^{10}-\frac{1}{2}
\left(\frac{\sigma_i(s)}{z}\right)^4 \right] 
\label{eq-Vwall-CG}
\end{equation}
where $\varepsilon_w^{\rm (CG)}$ is the interaction strength between soft spheres and the walls, $z$ and $L_z-z$ are the vertical distances from the two walls, respectively. 

\begin{figure*}[t!]
\begin{center}
(a)\includegraphics[width=0.32\textwidth,angle=270]{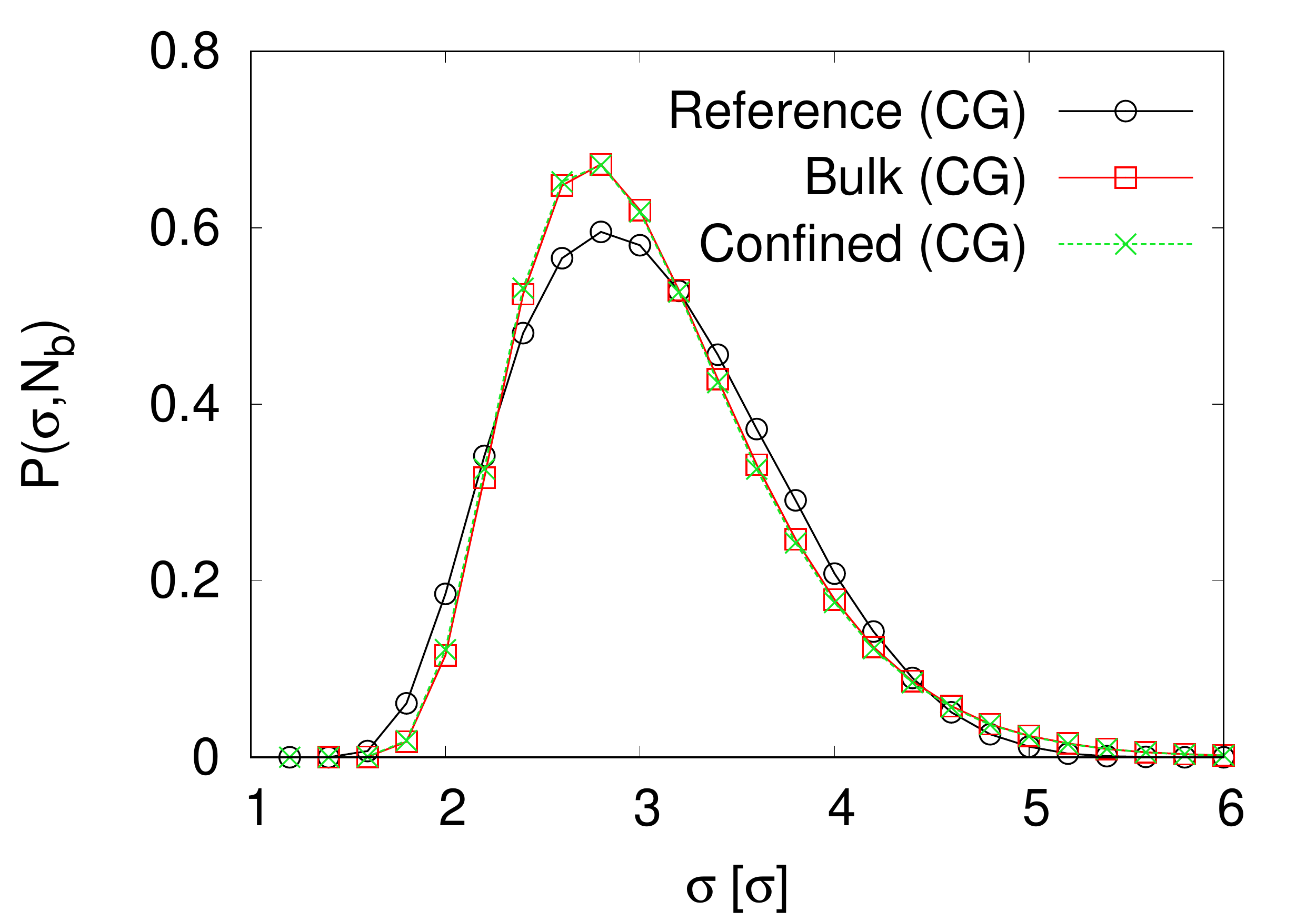} \hspace{0.1cm}
(b)\includegraphics[width=0.32\textwidth,angle=270]{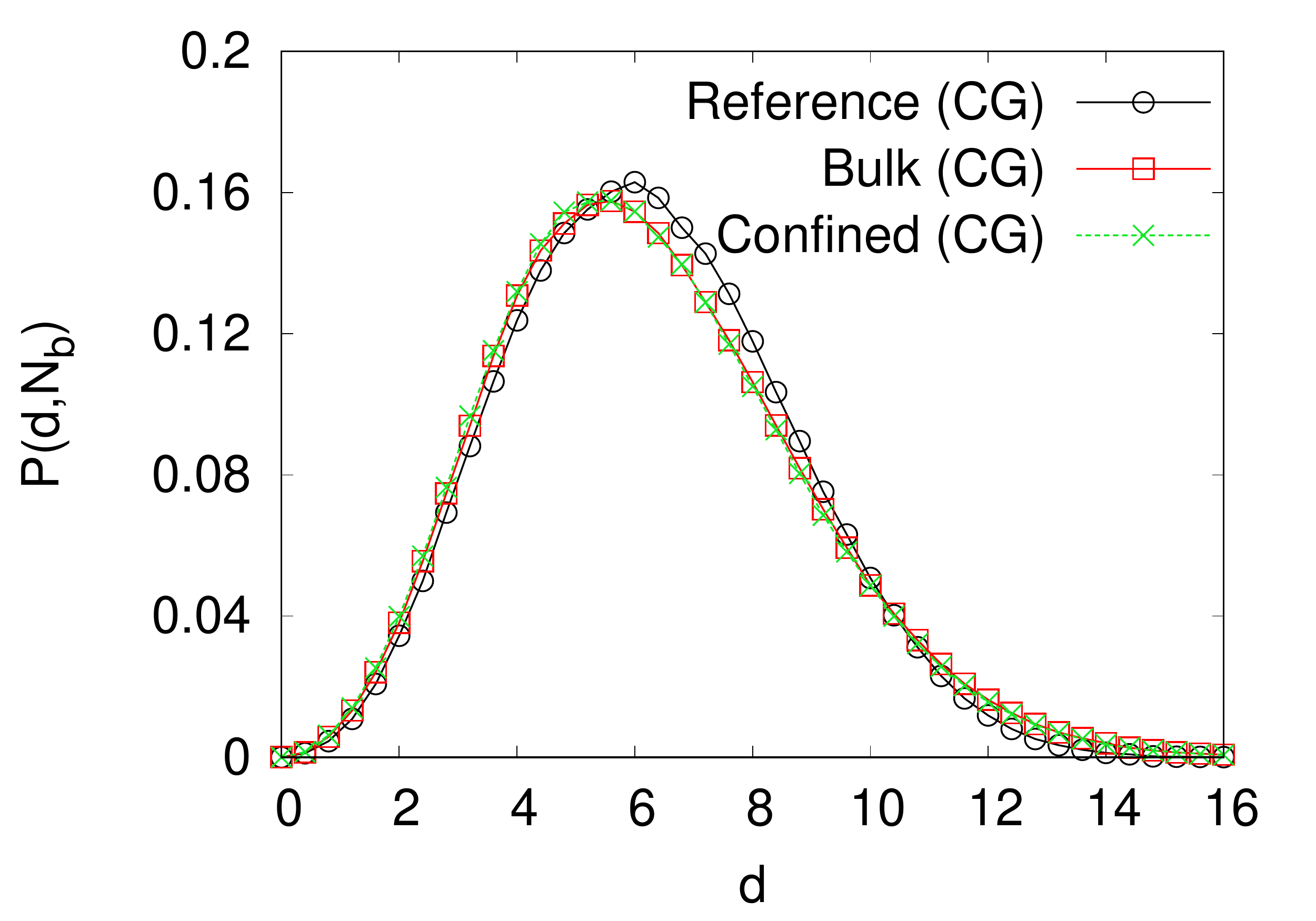}\\
(c)\includegraphics[width=0.32\textwidth,angle=270]{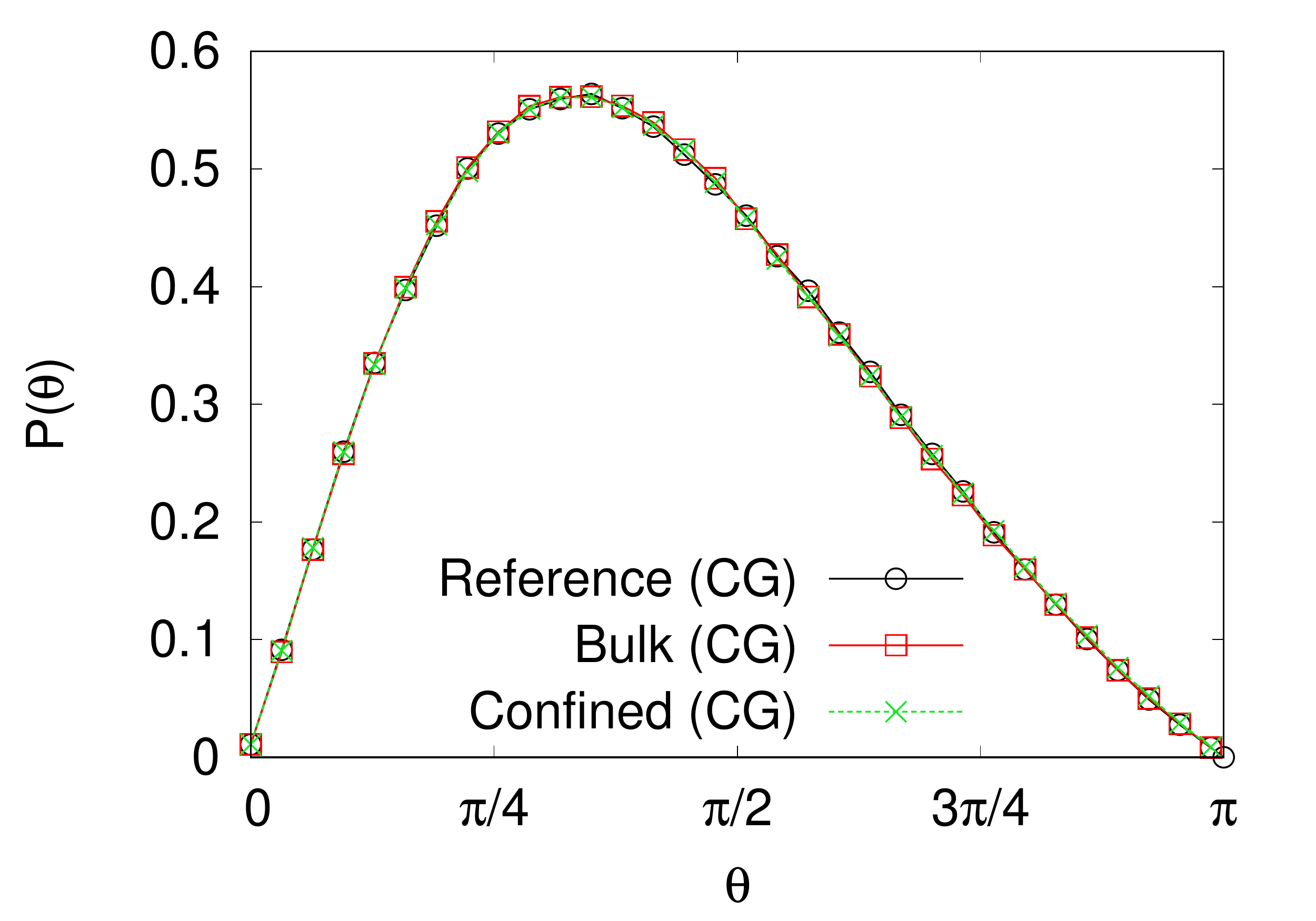}\\
\caption{Probability distributions of radius $\sigma$ of soft spheres, $P(\sigma)$ (a), bond length $d$ between two successive soft spheres, $P(d)$ (b), and the bond angle $\theta$ between two successive bonds, $P(\theta)$ (c), for a fully equilibrated bulk CG melt, and a confined CG melt ($n_c=1000$, $N_{\rm CG}=80$). Data for the reference systems ($n_c=1000$, $N=2000$) in a CG representation are also shown for comparison. Data are averaged over $30$ (bulk) and $60$ (confined melts) independent configurations.}
\label{fig-psigma-meltCG}
\end{center}
\end{figure*}

\begin{figure*}[t!]
\begin{center}
(a)\includegraphics[width=0.32\textwidth,angle=270]{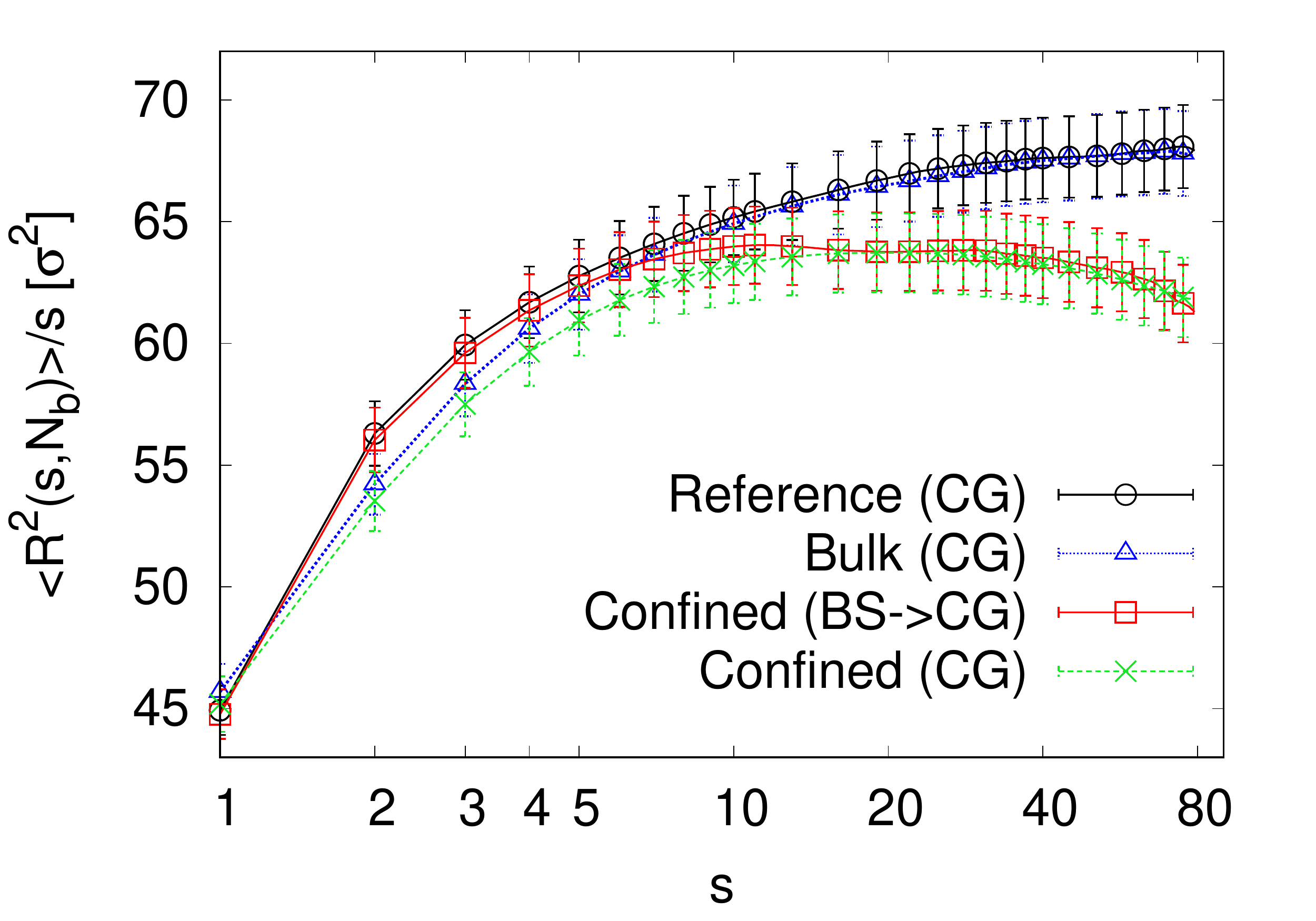} \hspace{0.1cm}
(b)\includegraphics[width=0.32\textwidth,angle=270]{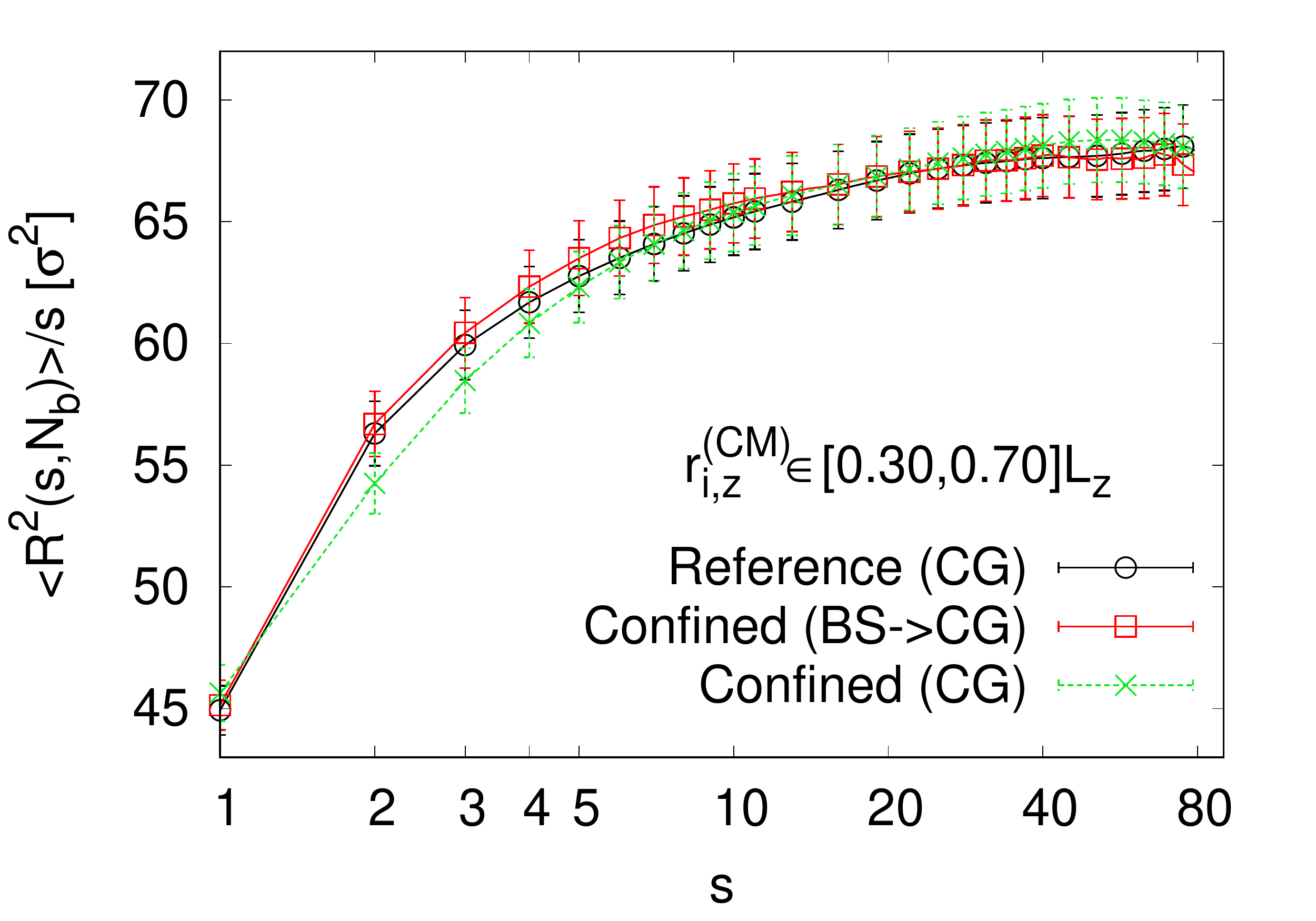}\\
(c)\includegraphics[width=0.32\textwidth,angle=270]{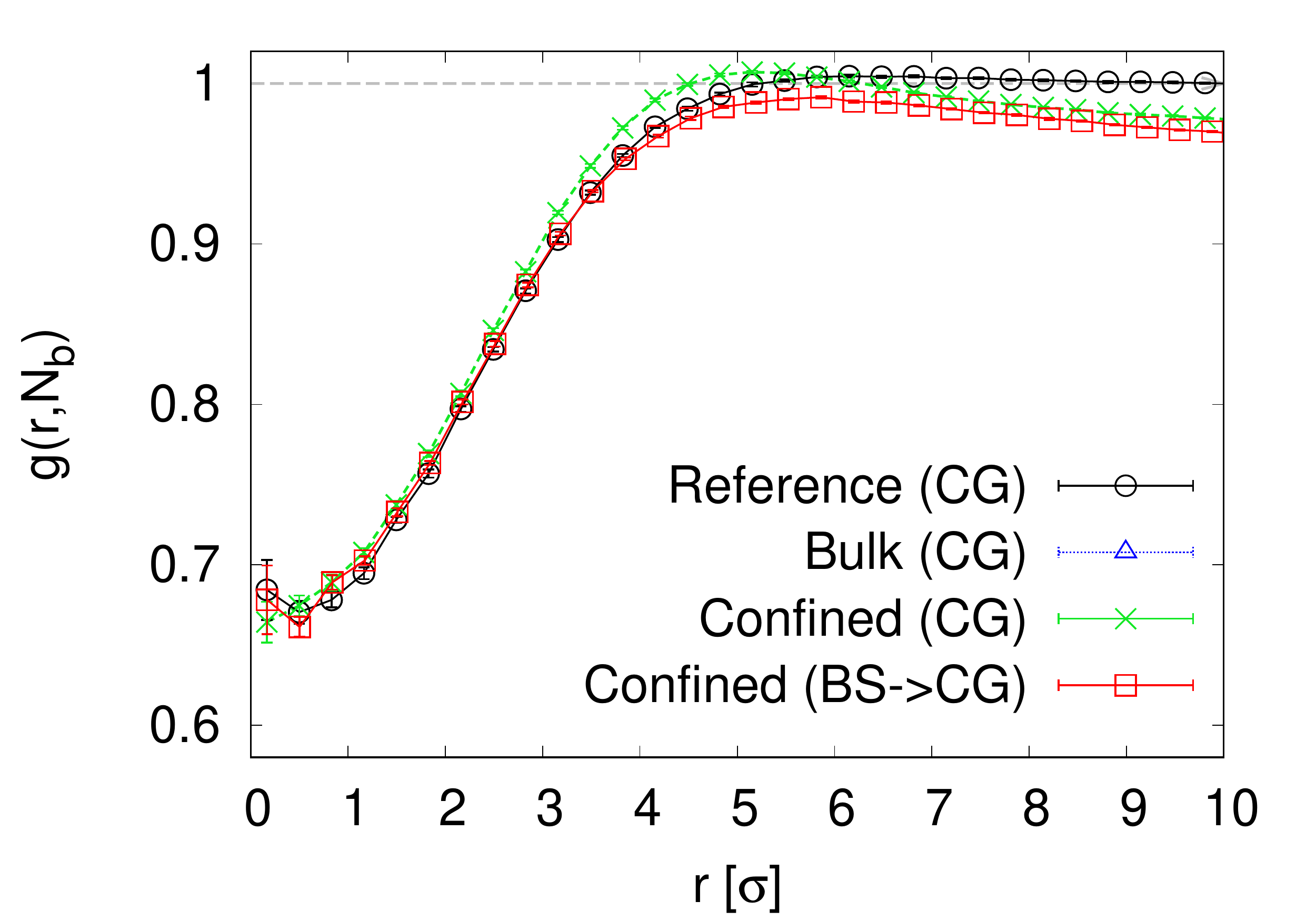} \hspace{0.1cm}
(d)\includegraphics[width=0.32\textwidth,angle=270]{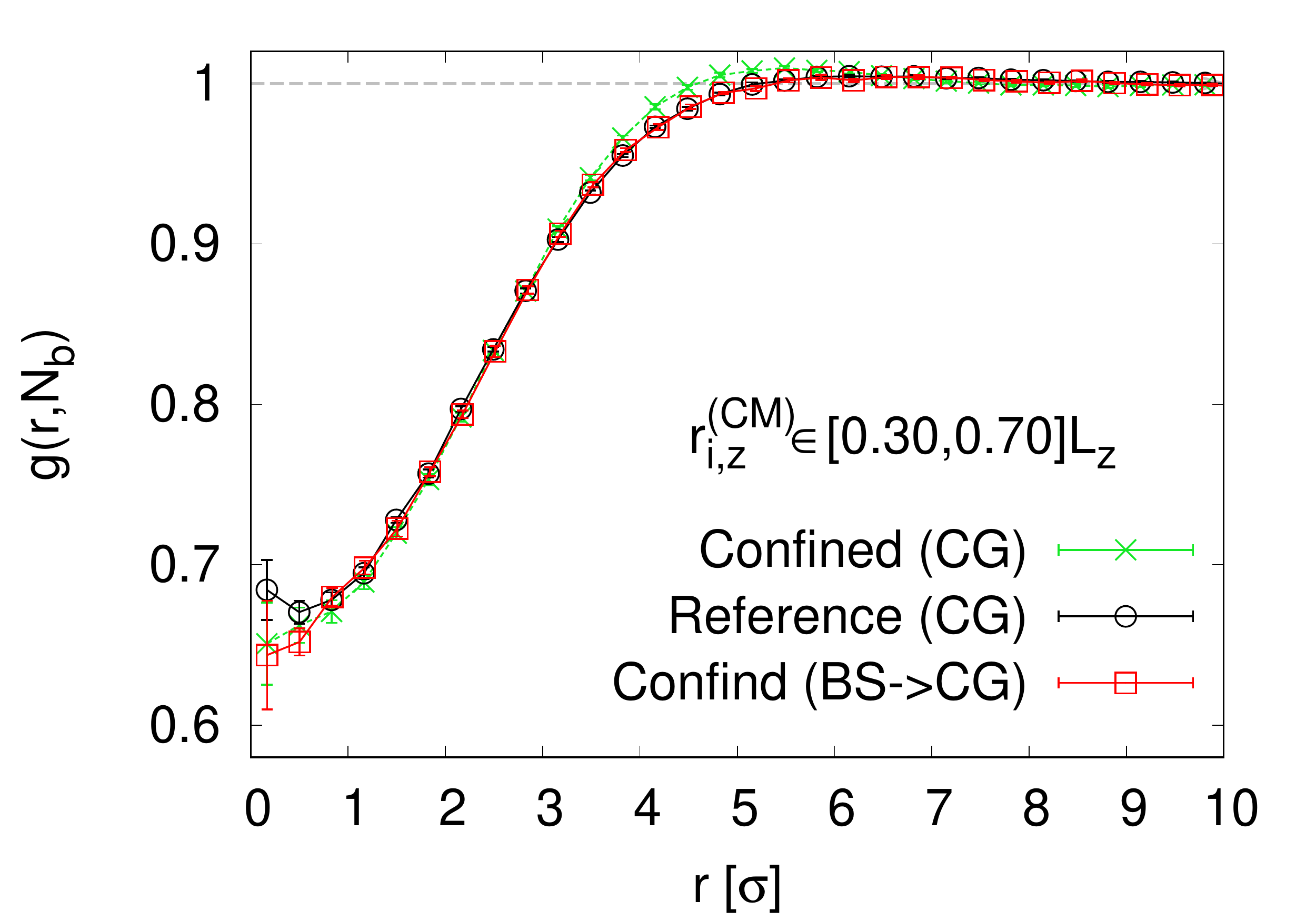}\\
\caption{(a)(b) Rescaled mean square internal distance, $\langle R^2(s,N_b) \rangle / s$, plotted as a function of $s$. (c)(d) Pair distribution of sphere pairs, $g(r,N_b)$, plotted as a function of $r$. Data for a bulk CG melt, a confined CG melt, a confined melt in a CG presentation are presented, as indicated. In (a)(c), all $n_c=1000$ chains are considered while only chains having their CMs in the interval $[0.3L_z, 0.7L_z]$ are counted in (b)(d). Data for the reference systems in a CG representation are also shown for comparison.}
\label{fig-R2s-wall}
\end{center}
\end{figure*}

For the parameterization of the soft-sphere CG model, we take $15$ independent and fully equilibrated bulk polymer melts of bead-spring polymer chains with $k_\theta=1.5\epsilon$ obtained from the previous works~\cite{Zhang2014,Moreira2015,Hsu2016} as our reference systems. Using Eqs.~(\ref{eq-mapping-cm}), (\ref{eq-mapping-rg}) with $N_b=25$, each melt containing $n_c=1000$ bead-spring chains of $N=2000=N_{\rm CG}N_b$ monomers in a cubic simulation box of size $V=L^3$ ($L\approx 133 \sigma$ with periodic boundary conditions in $x$-, $y$-, and $z$- directions) at the melt density $\rho_0=0.85\sigma^{-3}$ is mapped into a CG melt containing $n_c=1000$ soft-sphere chains of $N_{\rm CG}=80$ spheres at the CG melt density $\rho_0^{\rm (CG)}(N_{\rm CG})=\rho_0N_{\rm CG}/N=0.034\sigma^{-3}$. The parameters $a_1=6.3444\times 10^{-3}\sigma^6$, $a_2=4.1674\sigma^{-2}$, $k_{\rm bend}=1.3229\epsilon$, $a_3=22.0\epsilon \sigma^3$, $b_{\rm CG}=6.68\sigma \epsilon^{-1/2}$, and $\varepsilon_1=290.0\epsilon \sigma^{3}$ are determined~\cite{Vettorel2010,Zhang2013} such that  the average conformational properties of the reference melt systems in a CG representation are reproduced by fully equilibrated CG melts of soft-sphere chains. Quantitatively, the conformational properties are characterized by the probability distributions  of the radius of soft spheres, $P(\sigma,N_b)$, the bond length connecting two successive soft spheres, $P(d,N_b)$, and the bond angle between two successive bonds, $P(\theta,N_b)$, the average mean square internal distance between the $j$th soft sphere and the $(j+s)$th soft sphere along the identical chain, $\langle R^2(s,N_b) \rangle$, the pair distribution of all pairs of soft spheres, $g(r,N_b)$ (see Figs.~\ref{fig-psigma-meltCG}, \ref{fig-R2s-wall} discussed in the next section).

Since the excluded volume effect between $N_b$ monomers in each subchain is ignored in the {parameterization of the} soft-sphere approach and self-entanglements on this length scale is negligible ($N_b=25<N_e=28$) subchains behave as ideal chains {(alternative, one could include excluded volume for semi dilute solutions via a Flory term)}. In this case, we can simplify several steps of hierarchical backmapping~\cite{Zhang2014} to only one step of fine-graining to introduce microscopic details of subchains once a CG melt reaches its equilibrated state.

\section{Equilibration of soft-sphere chains in a confined CG melt}

In this section, we extend the application of the soft-sphere CG model for polymer  melts in bulk to polymer melts confined between two repulsive walls \{Eq.~(\ref{eq-Uwall-CG})\}, first focusing on the CG melt containing $n_c=1000$ chains of $N_{\rm CG}=80$ spheres at the CG bulk melt density $\rho_0^{\rm (CG)}=0.034\sigma^{-3}$. For the comparison to our reference systems, we set the distance between two walls compatible with the bulk melt. Thus, we locate two walls at $z=0\sigma$, and $z=L_z \approx 133 \sigma$ while keeping the periodic boundary conditions along the $x$- and $y$- directions with {the lateral linear dimensions} $L_x=L_y=133\sigma$. Of course, one can adjust $L_z$ and extend/reduce $L_x$, $L_y$ for keeping the bulk melt density as needed.

The initial configurations of soft-sphere chains in terms of $\{\sigma_i(s), d_i(s), \theta_i(s)\}$ are randomly generated according to their corresponding Boltzmann weights $\exp[-\beta U_{\rm sphere}^{\rm (CG)}(\sigma_i(s))]$,  $\exp[-\beta U_{\rm bond}^{\rm (CG)}(d_i(s))]$, and $\exp[-\beta U_{\rm ang}^{\rm (CG)}(\theta_i(s))]$, respectively, where $\beta=1/(k_BT)=1.0\epsilon^{-1}$. Additionally, we set $1\sigma<\sigma_i(s)<\sigma_{\rm max}=8\sigma$ and $0\sigma<d_i(s)<d_{\rm max}=21\sigma$ but restrict the coordinates of centers of spheres, $\vec{r}_i(s)$, satisfying the condition $\sigma_i(s)<\vec{r}_i(s)<(L_z-\sigma_i(s))$. It is computationally more efficient to perform Monte Carlo simulations to equilibrate confined CG melts. Similar to Ref.~\onlinecite{Zhang2013}, our simulation algorithm including three types of MC moves at each step is as follows: (i) For a local move, one of $n_cN_{\rm CG}$ spheres is randomly selected, e.g. the $s$th sphere in the $i$th chain, the sphere of radius $\sigma_i(s)$ at $\vec{r}_i(s)=(r_{i,x}(s), r_{i,y}(s), r_{i,z}(s))$ is allowed to move within the range $-\sigma_i(s)< \Delta r_{i,x}(s),\; \Delta r_{i,y}(s),\; \Delta r_{i,z}(s) < \sigma_i(s)$. The trial move is therefore accepted if $\exp[-\beta (\Delta U_{\rm nb}^{\rm (CG)}+\Delta U_{\rm bond}^{\rm (CG)}+\Delta U_{\rm ang}^{\rm (CG)} +\Delta U_{\rm wall}^{\rm (CG)})]>\eta$, where $\eta$ is a random number and $\eta \in[0,1)$. (ii) For a snake-slithering move, one end of $n_c$ chains is randomly selected, and $\sigma_i^{\rm (new)}(s)$, $d_i^{\rm (new)}(s)$, and $\theta_i^{\rm (new)}(s)$ of the selected sphere are randomly generated according to their corresponding Boltzmann weights, respectively. The trial move is accepted if $\exp[-\beta (\Delta U_{\rm nb}^{\rm (CG)}+\Delta U_{\rm self}^{\rm (CG)}+\Delta U_{\rm wall}^{\rm (CG)})]>\eta$. A cut-off at $r_{\rm cut}^{\rm (CG)}=20\sigma$ for calculating the non-bonded interactions between two  different spheres, $U_{\rm nb}^{\rm (CG)}(r_i(s),\sigma_i(s),r_{i'}(s'),\sigma_{i'}(s'))$, is also  introduced~\cite{Vettorel2010} since the contributions for $r>20\sigma$ are negligible. Nevertheless, there is no influence on measurements of any physical observable while it speeds up the simulations by a factor of four. Applying a linked-cell algorithm with the cell size $L_c=2.66\sigma$ ($L_x/L_c=L_y/L_c$ is very close to an integer), smaller than the cut-off value $r_{\rm cut}^{\rm (CG)}$, speeds up the simulation even more by an additional factor of {2.5, i.e. all together it speeds up by a factor of ten.} It takes about $10$ hours CPU time on an Intel 3.60GHz PC for a confined CG melt to reach its equilibrated state (after $2\times 10^7$ MC steps are performed). The acceptance ratio is about 73\% for a trial change of sphere size, 45\% for a local move, and 41\% for a snake-slithering move.

Choosing the wall strength $\varepsilon_{w}^{\rm (CG)}\approx {\cal O}(0.1-1)\epsilon$, there is no detectable influence on the probability distributions $P(\sigma,N_b)$, $P(d,N_b)$, and $P(\theta)$ comparing to that for an equilibrated CG melt in bulk as shown in Fig.~\ref{fig-psigma-meltCG}. Since soft spheres are allowed to penetrate each other in CG melts, we observe that the distributions $P(\sigma,N_b)$ and $P(d,N_b)$ are slightly narrower for both {equilibrated} CG melts in bulk  and in confinement than that for the reference systems while the average values of radius and bond length remain the same. We have also compared the estimates of $\langle R^2(s,N_b)\rangle$ and $g(r,N_b)$ for {an equilibrated} confined CG melt to an equilibrated CG melt in bulk and the reference data in Fig.~\ref{fig-R2s-wall}. The curve of $\langle R^2(s,N_b) \rangle$ taken the average over all $n_c=1000$ chains for a confined CG melt deviates from the bulk behavior for $s>10$ due to the confinement effect while for $s<6$,  $\langle R^2(s,N_b) \rangle$ is a bit smaller compared to the bulk value. It is due to the artifact of  the soft-sphere CG model, where the excluded volume effect within the size of spheres is not considered. After monomers are reinserted into soft-sphere chains, local excluded volume and the corresponding correlation hole effect automatically correct for these deviations. However, the discrepancy for $s<6$ is still within fluctuations observed in bulk. When monomers are reinserted into soft-sphere chains, the estimate of $g(r,N_b)$ starts to increase at $r \approx 3\sigma$, and then decrease at $r\approx 5 \sigma$. It indicates that near the walls, the distance between any two spheres decreases due to the confinement effect.

\begin{figure*}[t!]
\begin{center}
(a)\includegraphics[width=0.32\textwidth,angle=270]{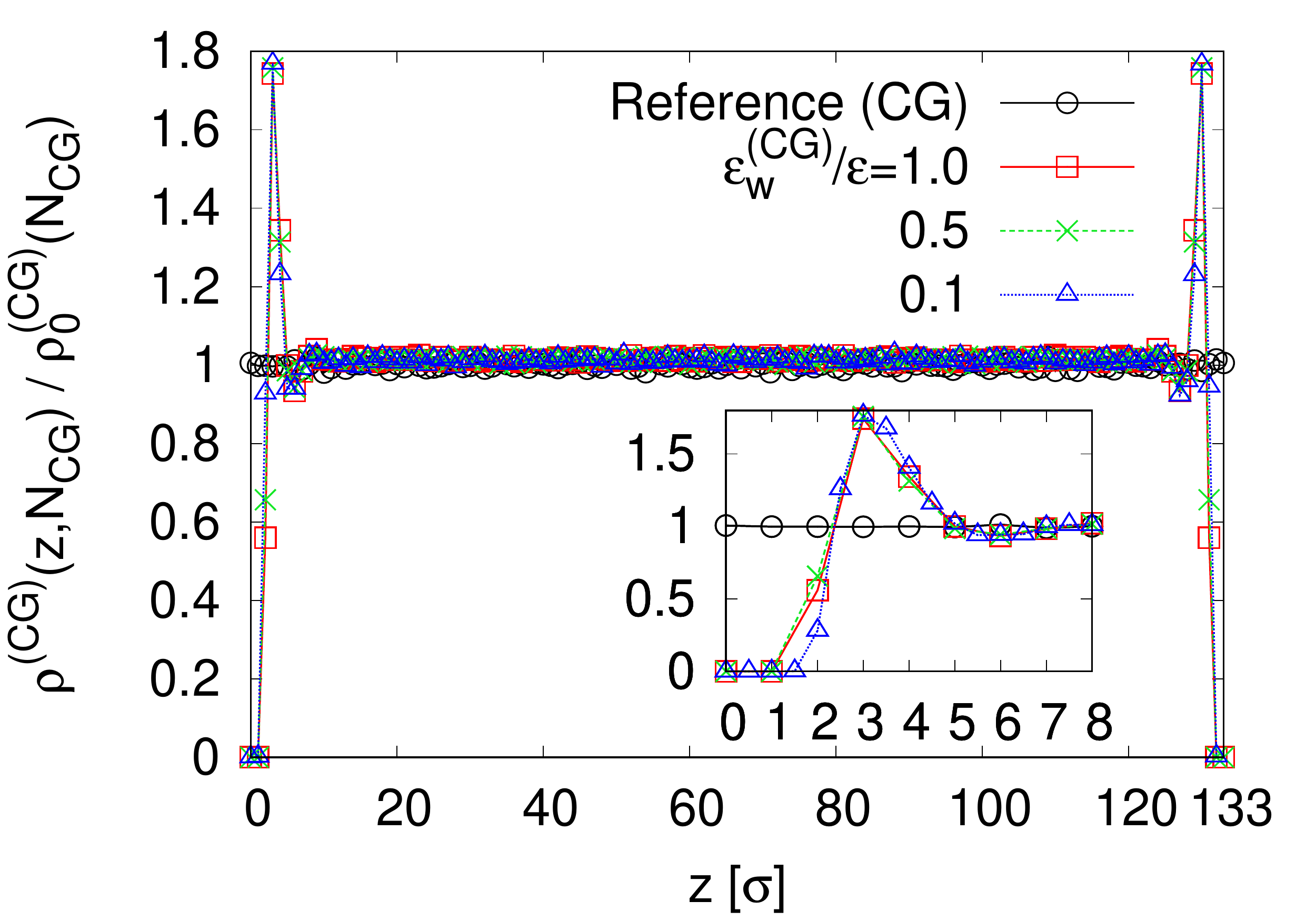} \hspace{0.1cm}
(b)\includegraphics[width=0.32\textwidth,angle=270]{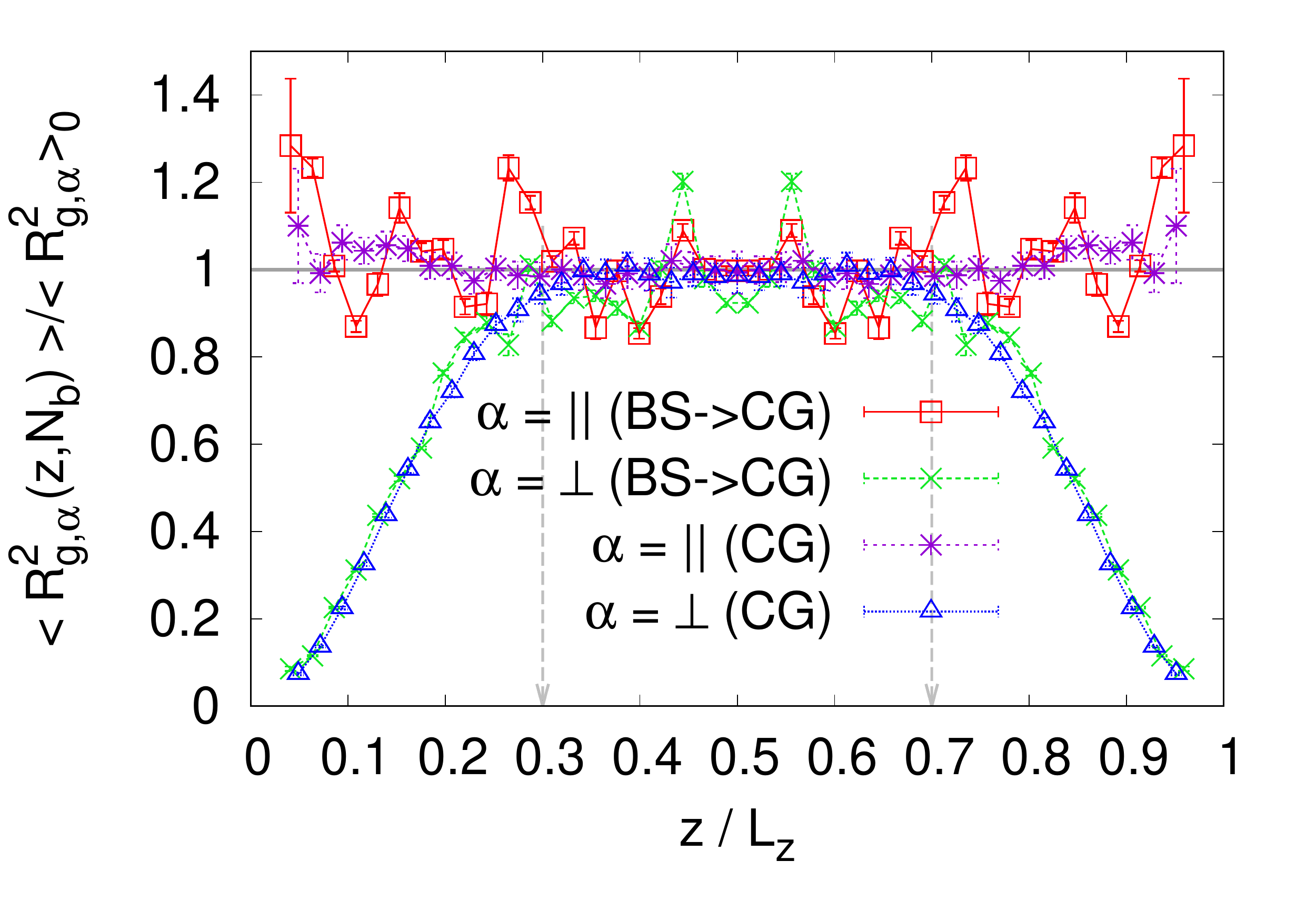}
\caption{(a) Soft-sphere density profile rescaled by the CG bulk melt density, $\rho^{\rm (CG)}(z,N_{\rm CG})/\rho_0^{\rm (CG)}(N_{\rm CG})$, plotted as a function of $z$ for fully equilibrated confined {CG} melts ($n_c=1000$, $N_{\rm CG}=80$, $N_b=25$) with three different values of $\varepsilon_{w}^{\rm (CG)}$, as indicated.
(b) Two components of the rescaled mean square radius of gyration in the  directions parallel ($||$) and perpendicular ($\perp$) to the walls,  $\langle R^2_{g,\alpha}(z,N_b) \rangle/\langle R^2_{g,\alpha} \rangle_0$, plotted versus the rescaled distance of the CM of the chains with bin size $3\sigma$ from the walls, $z/L_z$, including error bars for a confined CG melt with $\varepsilon_{w}^{\rm (CG)}=0.5\epsilon$ and $L_z=133\sigma$. In (a), Data for the reference systems in a CG representation are also shown for comparison. In (b), data for a confined melt of ($n_c=1000$, $N=2000$) based on the bead-spring model with $\varepsilon_{w}=0.005\epsilon$ and $L_z=134\sigma$ in a CG representation (BS $\rightarrow$ CG) are also shown for comparison.}
\label{fig-rhoz-wall}
\end{center}
\end{figure*}

To investigate the confinement effect on packing and conformations of a polymer melt, 
we determine the soft-sphere density profile between two walls as follows,
\begin{equation}
   \rho^{\rm (CG)}(z,N_b)=\frac{1}{L_x L_y}\sum_{i=1}^{n_c} \sum_{s=1}^{N_{\rm CG}}
  \delta(z_i(s)-z) \, .
\end{equation}
Fig.~\ref{fig-rhoz-wall} shows that the soft-sphere density profiles with bin size $1\sigma$ for three different values of the interaction strength between the soft spheres and walls, $\varepsilon_{w}^{\rm (CG)}/\epsilon=1.0$, $0.5$, and $0.1$ are the same within small fluctuation. The bulk melt density persists, i.e., $\rho^{\rm (CG))}(z,N_b) =\rho_0^{(\rm CG)}$, between $z=5$ and $z=L_z-5$. $\rho^{\rm (CG)}(z,N_b)$ increases and reaches a maximum value at $z=3\sigma$, and then approaches zero next to the walls. It indicates that the confinement effect is weak for spheres sitting in the middle regime between two walls.  The change of $\rho^{\rm (CG)}(z,N_b)$ is related to $P(\sigma,N_b)$ since $P(\sigma,N_b)$  has its maximum at $z=\langle \sigma_i(s) \rangle \approx 3.06\sigma$  (see Fig.~\ref{fig-psigma-meltCG}). The two components of the mean square radius of gyration depending on the $z$-component of CMs of chains are defined as follows,
\begin{equation}
    \langle R_{g,\alpha}^2(z,N_b) \rangle = 
\frac{\sum_{i=1}^{n_c} \sum_{s=1}^{N_{\rm CG}}
(\vec{r}_{i,\alpha}(s) - \vec{r}^{\rm (CM)}_{i,\alpha})^2 \delta(\vec{r}_{i,z}^{\rm (CM)}-z)}
{N_{\rm CG}\sum_{i=1}^{n_c}\delta(\vec{r}^{\rm (CM)}_{i}-z)}
\label{eq-Rg}
\end{equation}
where $\vec{r}_{i,\alpha=||}(s)=\vec{r}_{i,x}(s)+\vec{r}_{i,y}(s)$ and  $\vec{r}_{i,\alpha=\perp}(s)=\vec{r}_{i,z}(s)$. Fig.~\ref{fig-rhoz-wall}b shows that the linear dimensions of confined chains having their CM in the regime $0.3L_z\le \vec{r}^{\rm (CM)}_{i,z} \le 0.7L_z$ are the same as in a bulk melt {within fluctuation}. For polymer chains of $N=2000$ monomers in a bulk melt, the mean square radius of gyration $\langle R_g^2 \rangle_0=\langle R_{g,||}^2 \rangle + \langle R_{g,\perp}^2 \rangle \approx 909\sigma^2$. With decreasing the distance $z$ from the walls, $\langle R_{g,||}^2(z,N_b) \rangle$ increases moderately {with larger fluctuations} while $\langle R^2_{g,\perp}(z,N_b) \rangle$ decreases gradually even already in the regime where the monomer density $\rho^{\rm (CG)}(z,N_{\rm CG}) \approx \rho_0^{\rm (CG)}$. 
{Note that none of chains having their CM next to the wall}.
From the results shown in Fig.~\ref{fig-rhoz-wall}b, we should expect that $\langle R^2(s,N_b) \rangle$ and $g(r,N_b)$ follow the bulk behavior if we count those chains sitting in the middle regime ($0.3\le \vec{r}_{i,z}^{\rm (CM)} \le 0.7$) between two walls. It is indeed seen in Fig.~\ref{fig-R2s-wall}b,d. {Similar behavior has been observed for shorter chains confined between two walls~\cite{Pakula1991,Aoyagi2001,Cavallo2005,Sarabadani2014}.}

\section{Equilibrating bead-spring chains in a confined melt}

\subsection{Backmapping procedure}
\label{Backmapping}

\begin{figure*}[t!]
\begin{center}
\includegraphics[width=0.85\textwidth,angle=0]{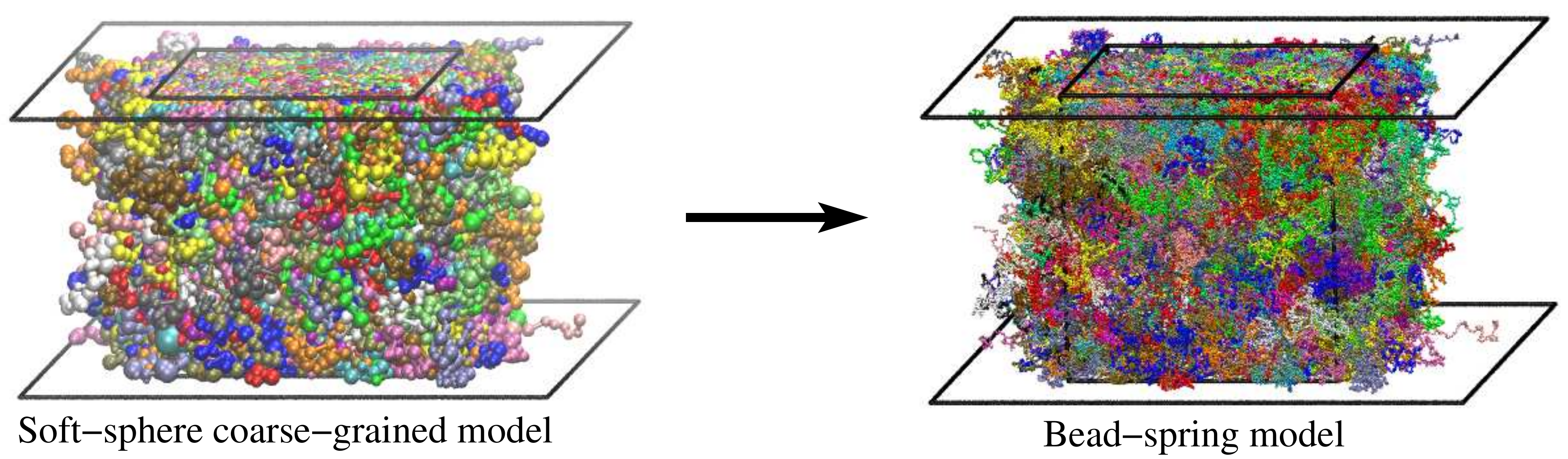}
\caption{Snapshots of the configuration of fully equilibrated CG melt containing $n_c=1000$ chains of $N_{\rm CG}=80$ soft spheres confined between two walls at the CG melt density $\rho_0^{\rm (CG)}(N_{\rm CG})=0.034\sigma^{-3}$, and its fine-graining configuration containing $n_c=1000$ chains of  $N=N_{\rm CG}N_b=2000$ monomers based on the bead-spring model at the melt density $\rho_0=0.85\sigma^{-3}$ obtained via the backmapping procedure (Sec.~\ref{Backmapping}). Here each soft sphere is represented by a subchain of $N_b=25$ monomers. The walls are placed at $z=0$ and $z=L_z$ where $L_z=133\sigma$ (left), and $134\sigma$ (right). The periodic boundary conditions are considered in the parallel directions to the walls, i.e., along the $x$- and $y$- directions, and the linear dimensions of walls are $L_x=L_y=133\sigma$.}
\label{fig-backmapping-wall}
\end{center}
\end{figure*}

After equilibration of the CG melt, we apply the similar backmapping strategy developed in Ref.~\onlinecite{Zhang2014} to reinsert the microscopic details of the bead-spring model described in Section~\ref{BSM} (See Fig.~\ref{fig-backmapping-wall}). In this strategy, two monomers along the chains are bonded via the FENE potential \{Eq.~(\ref{eq-UFENE})\} and the shifted LJ potential \{Eq.~(\ref{eq-ULJ})\}. The non-bonded and bond-bending interactions are excluded at this step. The confinement effect is introduced by the soft repulsive wall potential \{Eq.~(\ref{eq-Uwall})\}. Each soft-sphere CG chain of $N_{\rm CG}=80$ spheres is now replaced by a bead-spring chain of $N=2000$ monomers. To preserve the relationship between a soft sphere and a subchain of $N_b=25$ monomers given in Eqs.~(\ref{eq-mapping-cm}), (\ref{eq-mapping-rg}), two pseudopotentials~\cite{Zhang2014} for the $s$th soft sphere in the $i$th chain 
are implemented as follows,
\begin{equation}
    U_{\rm cm}(\vec{r}_i(s),\vec{R}_{\rm CM,i}(s))=k_{\rm cm} \left[\vec{r}_i(s)-\vec{R}_{{\rm CM},i}(s) \right ]^2
\end{equation}
and 
\begin{equation}
    U_g (\sigma_i,R_{g,i})= k_g \left[\sigma_i^2(s)-R_{g,i}^2(s)\right]^2
\end{equation}
where $k_{\rm cm}$ and $k_g$ determine the coupling strength. The forces derived from these two potentials can drive the center of mass and the radius of gyration of each subchain to the center and the radius of the corresponding soft sphere, respectively. Namely, each bead-spring chain is then sitting on top of its corresponding soft-sphere CG chain (see Fig.~\ref{fig-bsmsph}). During this backmapping procedure, it is more practical to perform MD simulations in the NVT ensemble with a weak coupling Langevin thermostat at $T=1.0\epsilon/k_B$ by setting the friction constant $\Gamma=0.5\tau^{-1}$. Choosing $k_{\rm cm}=50.0\epsilon$ and $k_g=5.0\epsilon$, the integration time step is set to $\Delta t=0.005\tau$. At this stage all $1000$ soft-sphere CG chains can be mapped into $1000$ bead-spring chains confined between two walls simultaneously since there is no interaction between different chains. Snapshots of the configurations of the fully equilibrated confined CG melt and the backmapped confined melt of bead-spring chains are shown in Fig.~\ref{fig-backmapping-wall}. Here the strength of the wall potential $U_{\rm wall}(z)$ is set to $\varepsilon_{w}=0.005\epsilon$. To keep the bulk melt density $\rho_0=0.85\sigma^{-3}$ in the middle regime between two walls (the weak confinement regime) in a microscopic representation, we set $L_z=134\sigma$ instead of $L_z=133\sigma$ taking the repulsive potential of the wall, which is a steep but smooth function, into account. The reinsertion MD time is about $80\tau$ and the CPU time is about $2.8$ hours on a single processor
in an Intel 3.60GHz PC.

\subsection{Equilibration procedure}
\label{Equilibration}  
\begin{figure*}[t!]
\begin{center}
(a)\includegraphics[width=0.32\textwidth,angle=270]{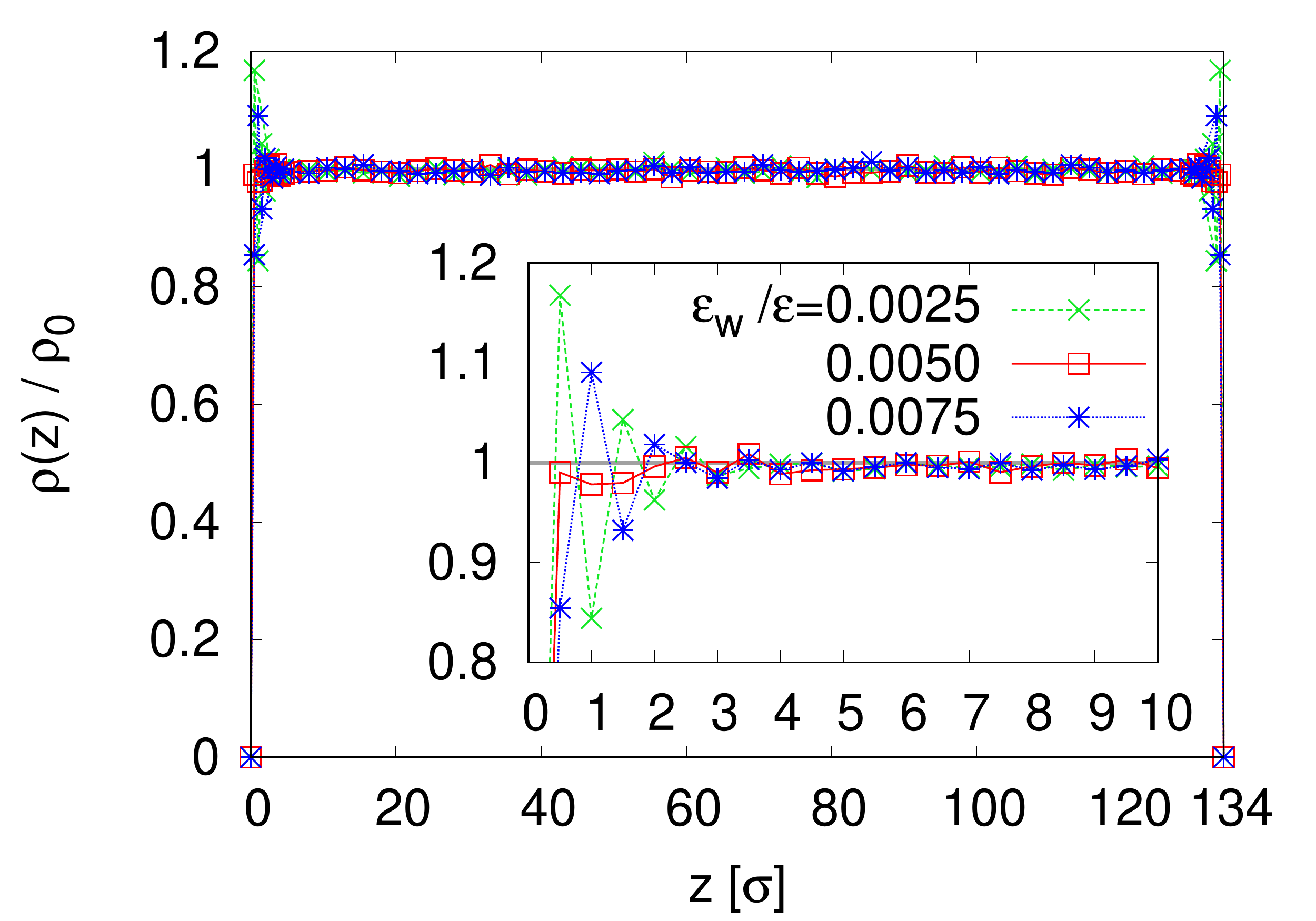} \hspace{0.1cm}
(b)\includegraphics[width=0.32\textwidth,angle=270]{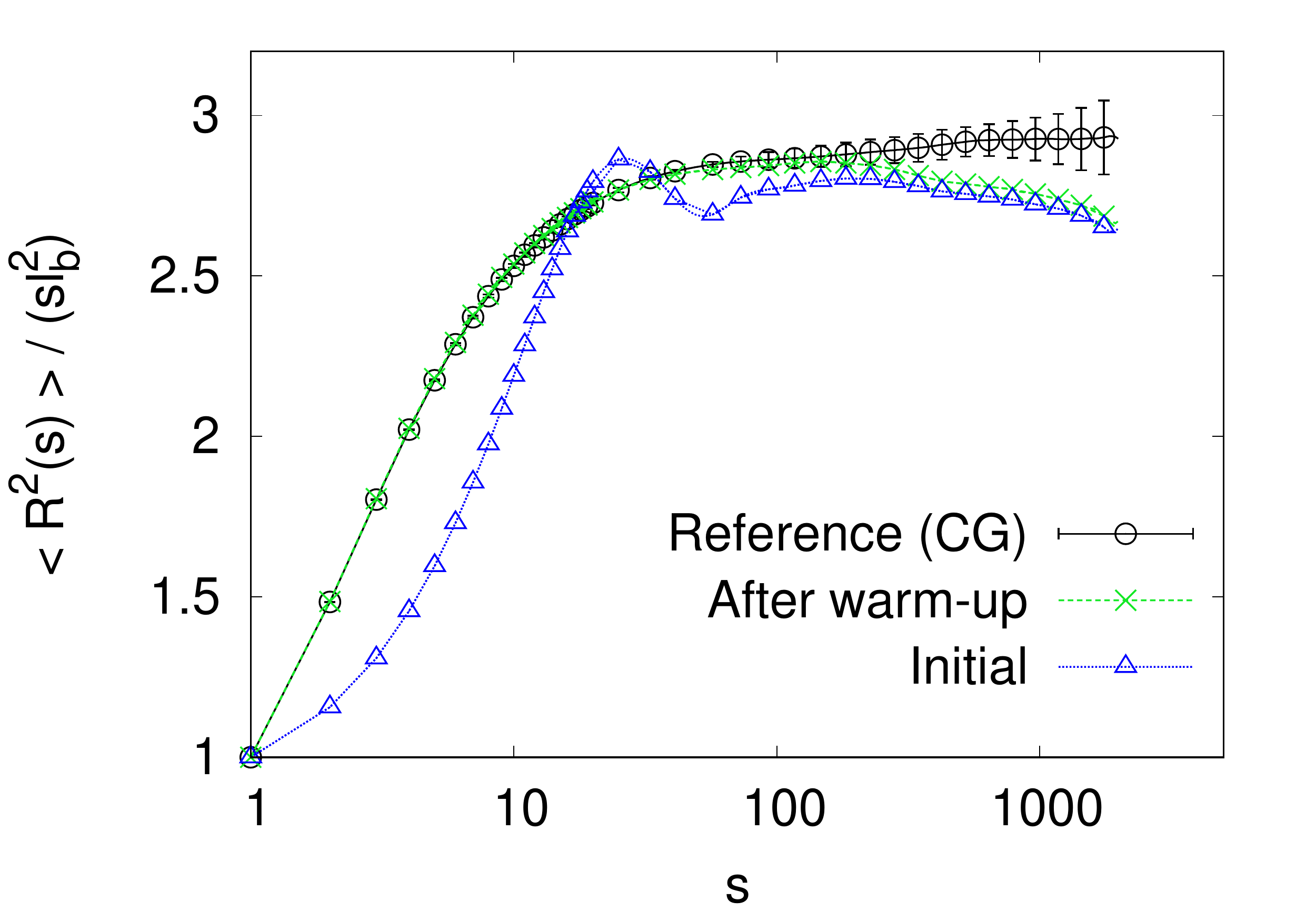}\\
(c)\includegraphics[width=0.32\textwidth,angle=270]{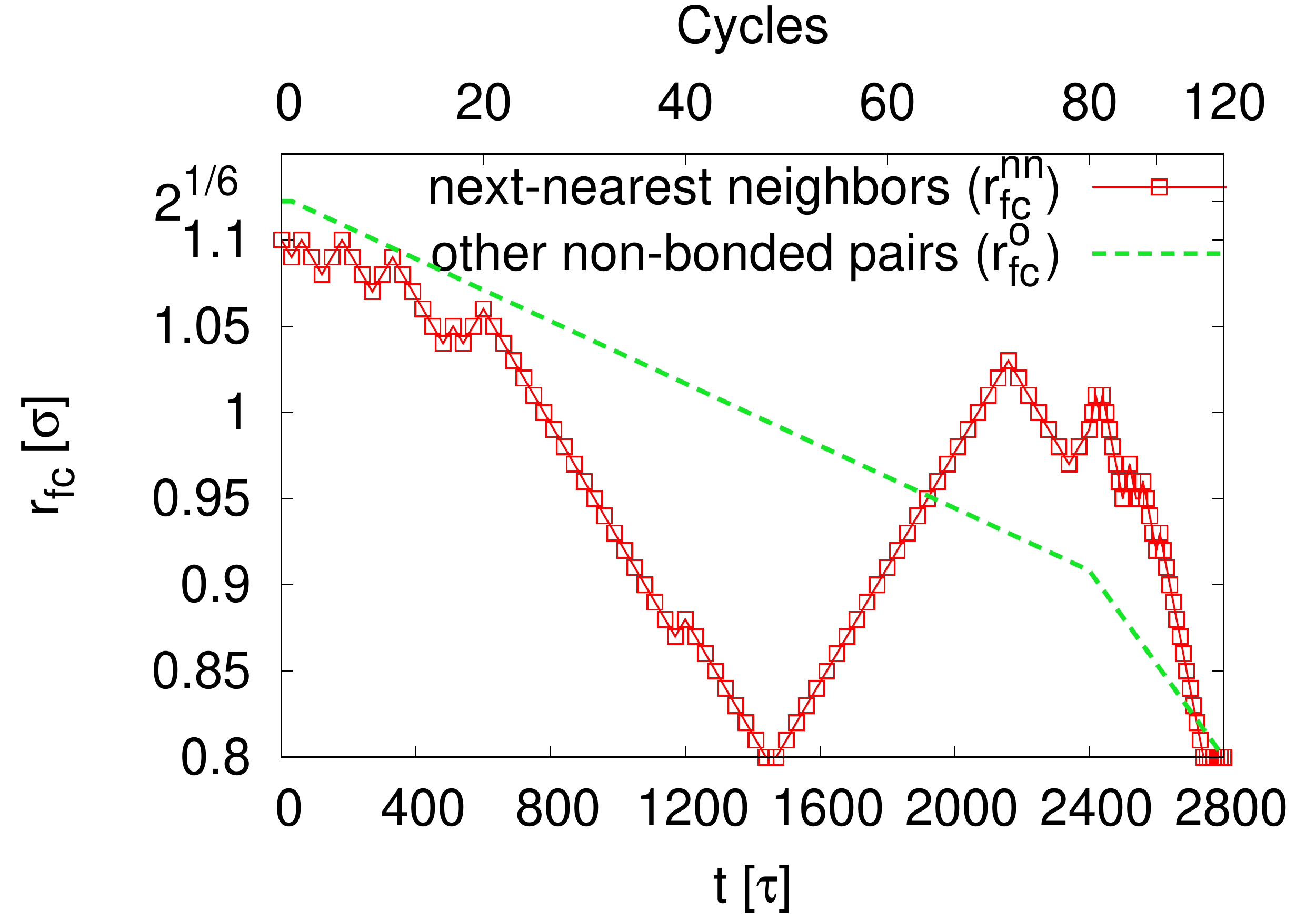}
\caption{(a) Monomer density profile rescaled by the bulk melt density $\rho(z)/\rho_0$, plotted as a function of $z$ for a polymer melt confined between two walls located at $z=0\sigma$ and $z=L_z=134\sigma$ in the intermediate stage of the warm-up procedure. (b) Rescaled mean square internal distance, $\langle R^2(s) \rangle / (sl_b^2)$, plotted versus $s$ for chains in a confined polymer melt right after the backmapping and warm-up procedures, $\ell_b=0.964\sigma$ being the mean square root of bond length. (c) Time series of the cut-off distance for the next-nearest neighboring pairs of monomers ($r_{\rm fc}^{\rm nn}$), and other non-bonded pairs ($r_{\rm fc}^{\rm o}$) for $U_{\rm FC-LJ}(r)$ in a warm-up procedure. In (a), several values of the strength $\varepsilon_w$ are chosen, as indicated, and only data for $z \le 10\sigma$ are shown in the inset. In (b), the master curve for the average behavior of the reference systems is also shown for comparison.}
\label{fig-Warmup-wall}
\end{center}
\end{figure*}

In the next step the full excluded volume interaction as listed in Section~\ref{BSM} has to be introduced. To avoid the ``explosion'' of the system due to the overlap of monomers, we have to switch on the excluded volume interactions between the non-bonded pairs of monomers in a quasi-static way (slow push-off)~\cite{Auhl2003, Moreira2015}. Therefore, the shifted LJ potential for each non-bonded pair of monomers at a distance $r$, \{Eq.~(\ref{eq-ULJ})\}, is first replaced by a force-capped LJ potential 
\begin{eqnarray}
    U_{\rm FC-LJ}(r)=\left\{\begin{array}{ll}
   (r-r_{\rm fc})U_{\rm LJ}^{'}(r_{\rm fc})+U_{LJ}(r_{\rm fc}) &  {\rm for} \, r\le r_{\rm fc}  \\
   U_{\rm LJ}(r)  & {\rm for} \, r > r_{\rm fc} 
  \end{array} \right .  
\end{eqnarray}
where $r_{\rm fc}$ is an adjustable cut-off distance in this warm-up procedure. 
{$r_{\rm fc}=r_{\rm fc}^{o}$ decreases monotonically at each cycle from $2^{1/6}\sigma$ to $0.8\sigma$ in the warm-up procedure
for the non-bonded pairs except the next-nearest neighboring (nn) pairs along identical chains.
For the nn pairs, $r_{\rm fc}=r_{\rm fc}^{\rm nn}$ is set to $1.1\sigma$ initially, and
tuned according to the following cost function~\cite{Moreira2015}} 
\begin{equation}
     C = \int_{s=20}^{s=50} ds \left [ \left(\frac{\langle R^2(s) \rangle}{s} \right)_{\textrm{master curve}}
 -\left(\frac{\langle R^2(s) \rangle}{s} \right)_{\textrm{current cycle}} \right ]
\label{eq-cost}
\end{equation}
at the end of each cycle in the warm-up procedure with the restriction that $0.8\sigma \le r_{\rm fc}^{\rm nn} \leq 1.1\sigma$. Master curve refers to fully equilibrated polymer melts in bulk (the reference systems). In this process the nn-excluded volume is adjusted according to the value of $C$, namely reduced (increase $r_{\rm fc}^{\rm nn}$ by $0.01\sigma$) if $C<0$ and enhanced (decrease $r_{\rm fc}^{\rm nn}$ by $0.01\sigma$) if $C>0$. If $|C|<0.0001\sigma^2$, $r_{\rm fc}^{\rm nn}$ remains unchanged. Thus, we have assumed that the mean square internal distance, $\langle R^2(s) \rangle$, for the confined chains in a melt finally should coincident with the master curve for $s<50$ at least. For a polymer melt under strong confinement effect, the bulk behavior may no longer be valid even for $s<50 \approx 2N_e$. However, it's more important to first correct chain distortion due to the absence of excluded volume interactions between monomers (see Fig.~\ref{fig-Warmup-wall}b). Once we remove the criterion of the cost function, confined chains will relax very fast on a short length scale dominated by both the entanglement and confinement effects. Note that this final equilibration step only affects subchain lengths of up to the order of $N_e$.

We perform MD simulations in the NVT ensemble with a Langevin thermostat at the temperature $T=1.0\epsilon/k_B$  using the package ESPResSo++~\cite{Espressopp,Espressopp20} for equilibrating a confined polymer melt containing $n_c=1000$ chains of $N=2000$ monomers under three procedures as follows: (a) In the warm-up procedure, $120$ cycles of $1.5 \times 10^{5}$ MD steps per cycle in the first $80$ cycles and $5 \times 10^4$ MD steps per cycle in the rest $40$ cycles are performed with a larger friction constant  $\Gamma=1.0\tau^{-1}$, and a small time step $\Delta t=0.0002\tau$. Differently from~\cite{Moreira2015}, more MD steps and a slower rate of decreasing the cut-off value of $r_{\rm fc}^o$ for non-bonded monomer pairs except nn pairs are required for the confined polymer melt system due to the competition between the excluded volume effect and the confinement effect. In the first $100$ cycles, $r_{\rm fc}^{\rm nn}$ is updated at each cycle associated with the  cost function. In the rest $20$ cycles, $r_{\rm fc}^{\rm nn}$ decreases by $0.01\sigma$ before reaching the minimum value $0.8\sigma$ (see Fig.~\ref{fig-Warmup-wall}a).  (b) In the relaxation procedure, the shifted LJ potential $U_{\rm LJ}(r)$ is restored. We first perform $10^5$ MD steps with $\Gamma=0.5\tau^{-1}$ and  $\Delta t=0.001\tau$, and then another $2 \times 10^6$ MD steps with $\Delta t=0.005\tau$ to ensure that the confined melt reaches its equilibrated state. Afterwards, we can set the time step to its standard value, $\Delta t =0.01\tau$ for the further  study of the confined polymer melt in equilibrium.

During warming up the confined polymer melt, the local monomer density profile $\rho(z)$ near the walls varies with the interaction strength $\varepsilon_w$ between monomers and walls as shown in Fig.~\ref{fig-Warmup-wall}a. For $\varepsilon_w=0.005\epsilon$, the bulk melt density $\rho_0$ is conserved all the way up to the wall, where the bin size is set to {$0.5\sigma$}. At the same time, after the confined polymer melt is warmed up, the curve of $\langle R^2(s) \rangle$ coincides with the master curve for $s<50$ as shown in  Fig.~\ref{fig-Warmup-wall}b. A typical variation of cut-off distances, $r_{\rm fc}^{\rm nn}$ and $r_{\rm fc}^{\rm o}$ for the non-bonded pairs of monomers in the warm-up procedure are  shown in Fig.~\ref{fig-Warmup-wall}c. At the end of the warm-up procedure, $U_{\rm FC-LJ}(r)$ approaches to $U_{\rm LJ}(r)$. Thus, there is no problem to switch back to $U_{\rm LJ}(r)$ for the further relaxation of the confined polymer melt.

\begin{figure*}[t!]
\begin{center}
(a)\includegraphics[width=0.32\textwidth,angle=270]{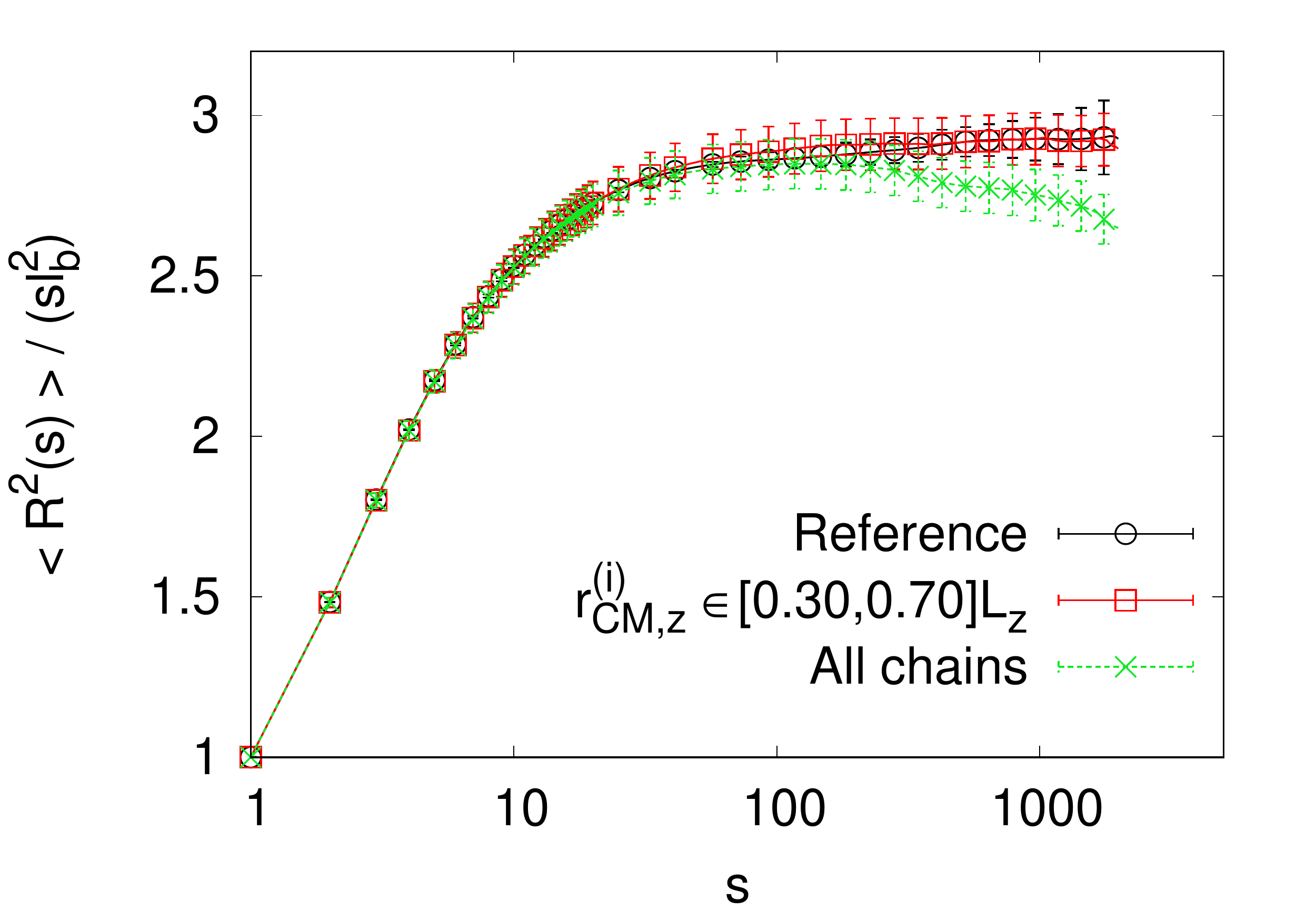}\hspace{0.1cm}
(b)\includegraphics[width=0.32\textwidth,angle=270]{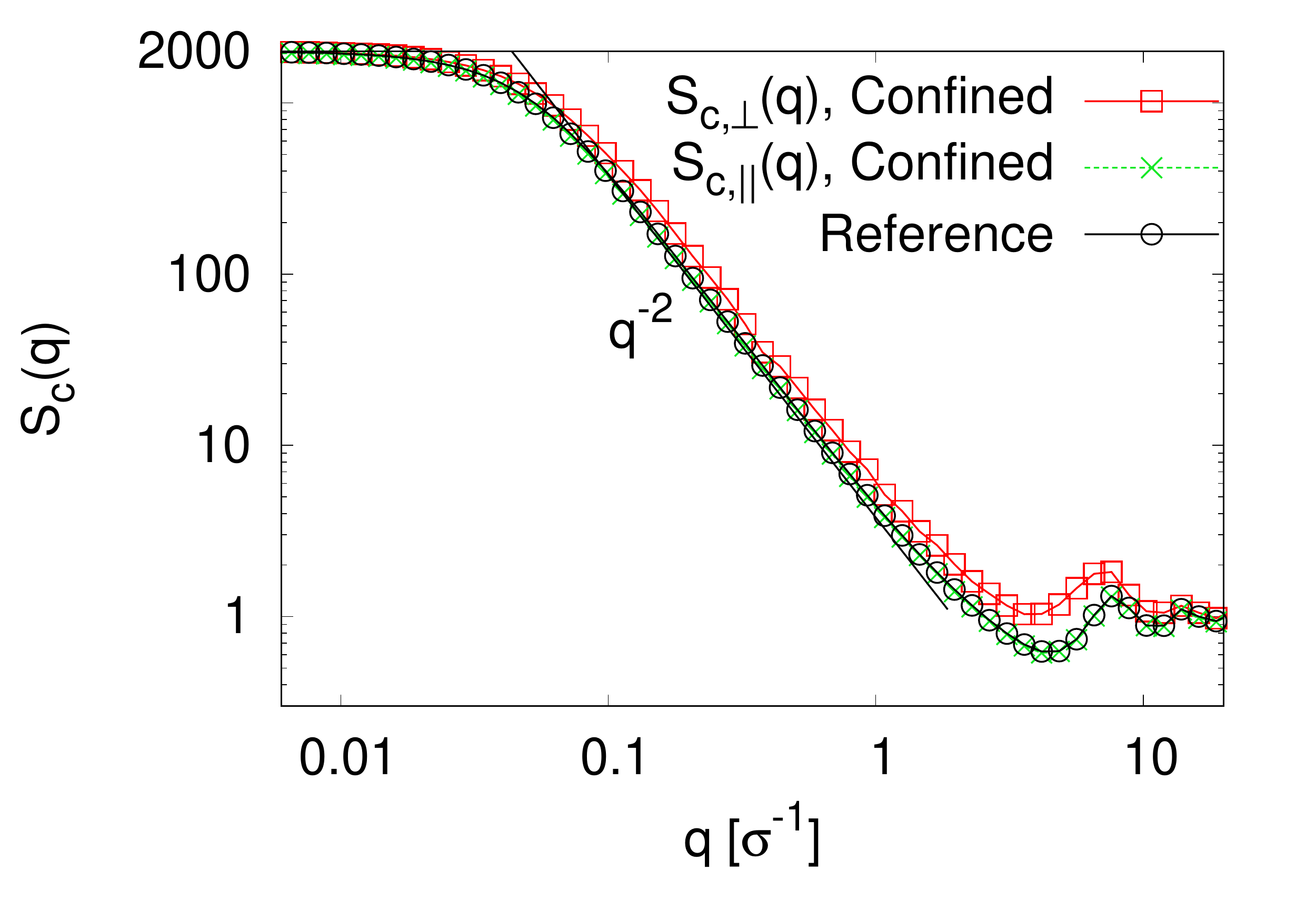}\\
(c)\includegraphics[width=0.32\textwidth,angle=270]{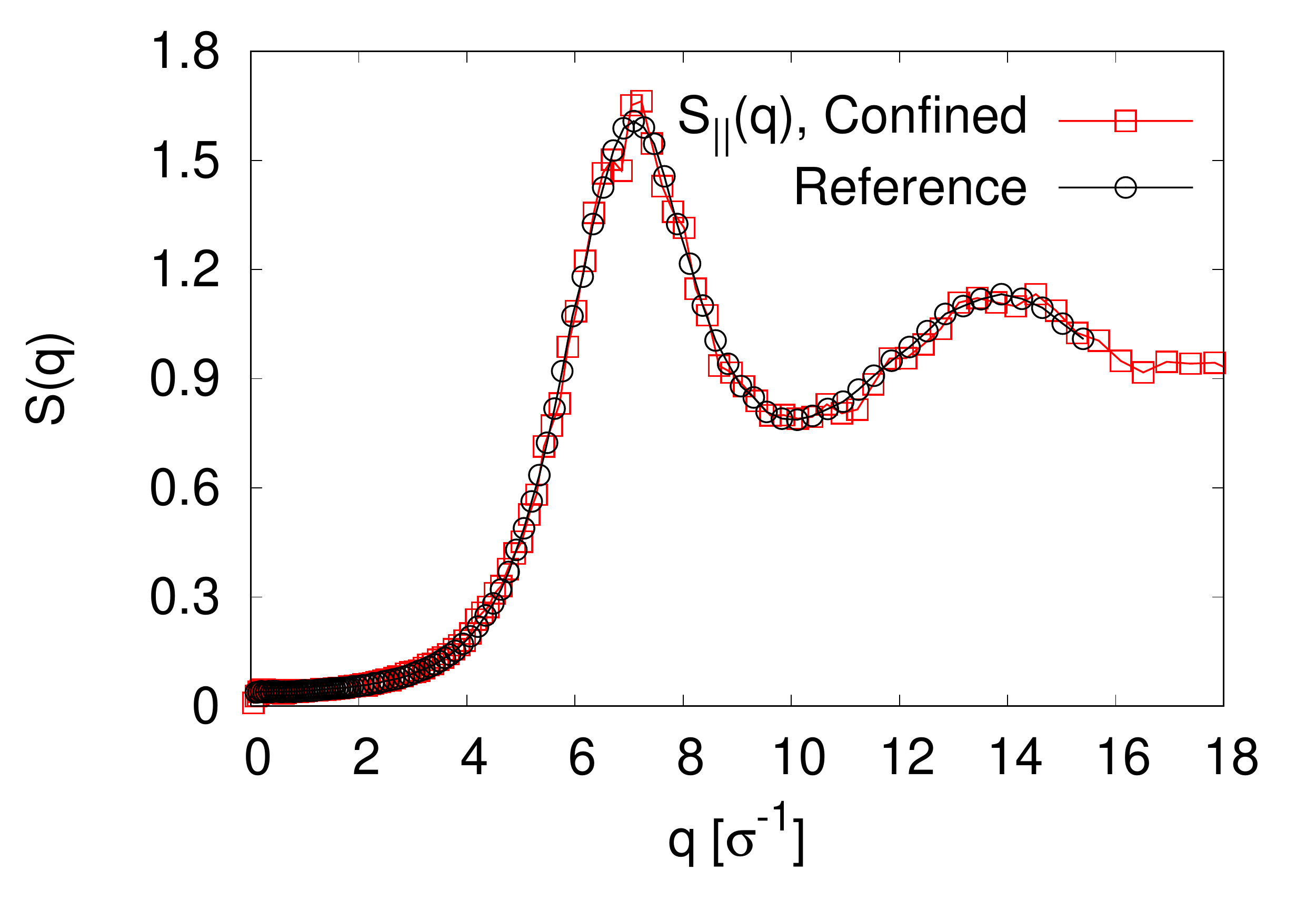}\hspace{0.1cm}
(d)\includegraphics[width=0.32\textwidth,angle=270]{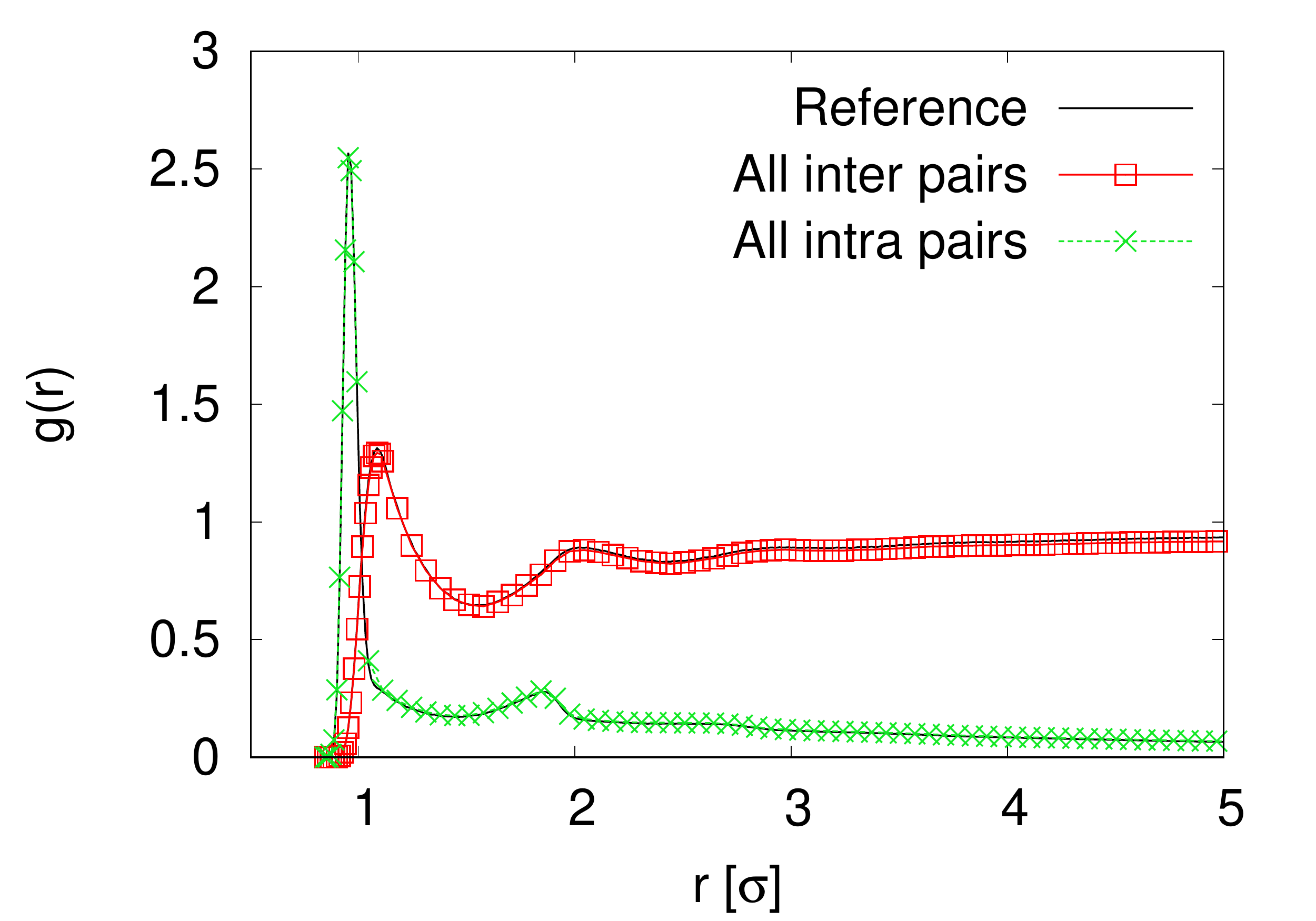}
\caption{(a) Rescaled mean square internal distance, $\langle R^2(s) \rangle / (sl_b^2)$, plotted as a function of $s$. (b) Two components of the single chain structure factors, $S_{c,||}(q)$ and $S_{c,\perp}(q)$, plotted versus $q$. (c) Collective structure factor $S(q)$ plotted versus $q$. (d) Pair distribution $g(r)$ plotted versus $r$. Data are for a fully equilibrated confined melt of bead-spring chains with the interaction strength $\varepsilon_w=0.005\epsilon$. In (b) the theoretically predicted slope $q^{-2}$ for a Gaussian coil is also shown for comparison. Data for the average behavior of fully equilibrated polymer melts in bulk (the reference systems) are also shown for comparison.}
\label{fig-eqbsm-wall}
\end{center}
\end{figure*}

First we examine the difference between the equilibrated confined melt and the reference bulk melt as shown in Fig.~\ref{fig-eqbsm-wall}. We compare the whole system in terms of the mean square internal distance $\langle R^2(s) \rangle$, the single chain structure factor $S_c(q)$, and the collective structure $S_{||}(q)$ in the direction parallel to the walls. 
We see that the curve of $\langle R^2(s) \rangle$ estimated only for those chains having their CMs satisfying  $0.3L_z\le \vec{r}_{i,z}^{\rm (CM)} \le 0.7L_z$ follows the master curve while the curve obtained by taking the average over all $1000$ chains starts to deviate from bulk behavior at $s=250$. To detect any anisotropy of chain conformations under confinement, we distinguish $S_{c,\perp}(q=q_z)$ where the wave vector $\vec{q}$ is oriented in the $z$-direction perpendicular to the wall, and $S_{c,||}(q=(q_x^2+q_y^2)^{1/2})$. $S_{c,||}(q)$ follows the same chain structure as in the bulk melt as shown in Fig.~\ref{fig-eqbsm-wall}b. The shift of {$S_{c,\perp}(q)$} toward to {a slightly larger} value of $q$ from the bulk curve also indicates that the estimate of {$\langle R^2_{g,\perp} \rangle$} for all $1000$ chains is smaller. Nevertheless, chains still behave as ideal chains. Comparison of the collective structure factor of the whole melt between the confined polymer and the bulk melt in the parallel direction to the walls, we see that there is no difference as the distance between two walls is compatible to the bulk melt (see Fig.~\ref{fig-eqbsm-wall}c). The local packing of monomers characterized by the pair distribution $g(r)$ for both inter and intra pairs of monomers is also in perfect agreement with the bulk melt as shown in Fig.~\ref{fig-eqbsm-wall}d. 

Finally, we go back to the outset and analyze the conformational properties of the confined equilibrated melt in a CG representation by mapping bead-spring chains to  soft-sphere chains using Eqs.~(\ref{eq-mapping-cm}), (\ref{eq-mapping-rg}). As shown in Fig.~\ref{fig-R2s-wall} and Fig.~\ref{fig-rhoz-wall}b, all results obtained by taking the average over $14$ configurations within $6\tau_e$ are consistent with the data obtained from the MC simulations of a confined CG melt within fluctuation. Since in the microscopic model, the excluded volume interactions between monomers are properly taken into account, the curves of $\langle R^2(s,N_b) \rangle$, $g(r,N_b)$ at short length scales do not deviate from the curves for the reference systems in a CG representation. This shows that the confined polymer melt indeed reaches the equilibrium on all length scales.

\section{Preparation of supported and free-standing films at zero pressure}

\begin{figure*}[t!]
\begin{center}
(a)\includegraphics[width=0.32\textwidth,angle=270]{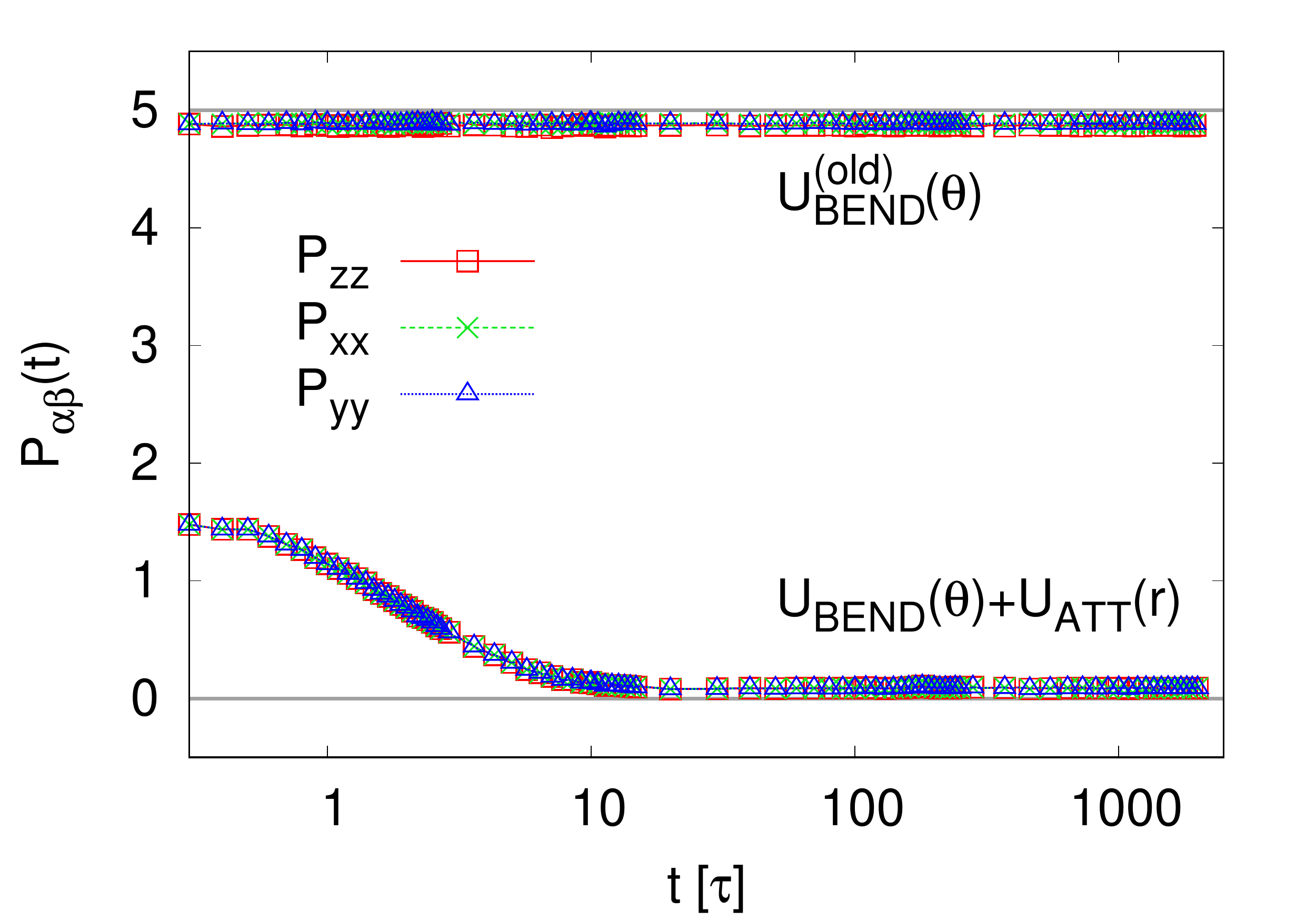}\hspace{0.1cm}
(b)\includegraphics[width=0.32\textwidth,angle=270]{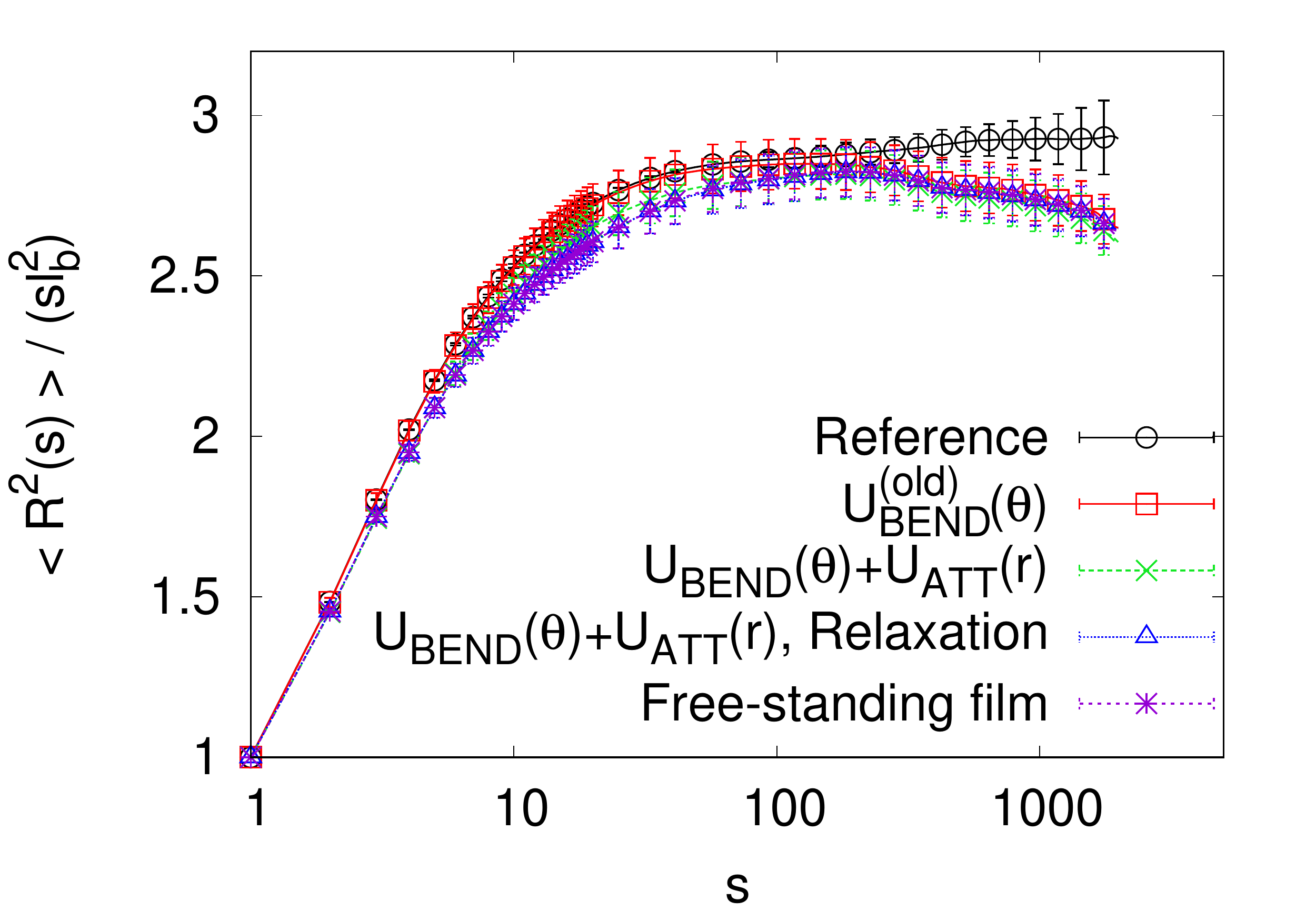}\\
(c)\includegraphics[width=0.32\textwidth,angle=270]{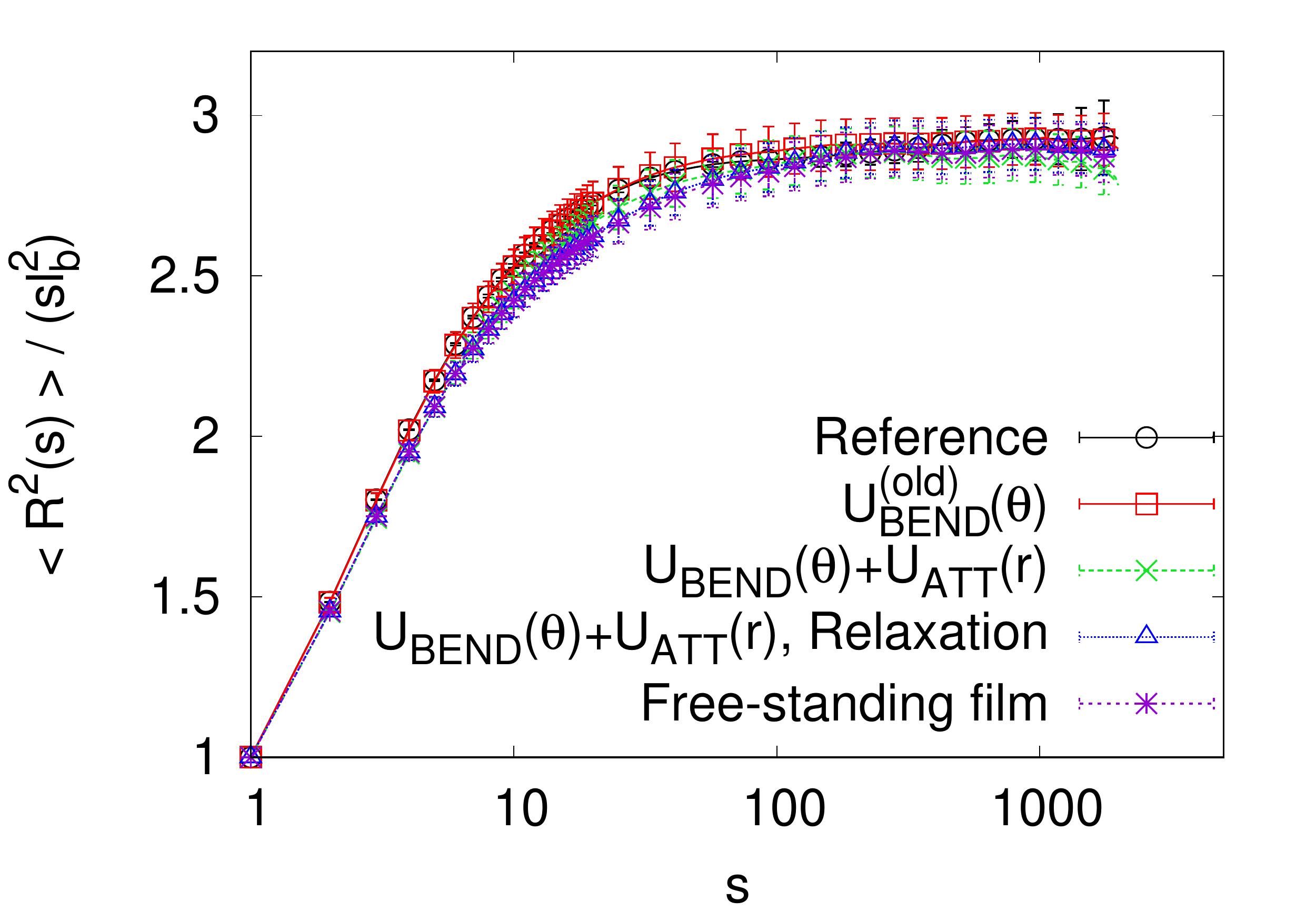}\hspace{0.1cm}
(d)\includegraphics[width=0.32\textwidth,angle=270]{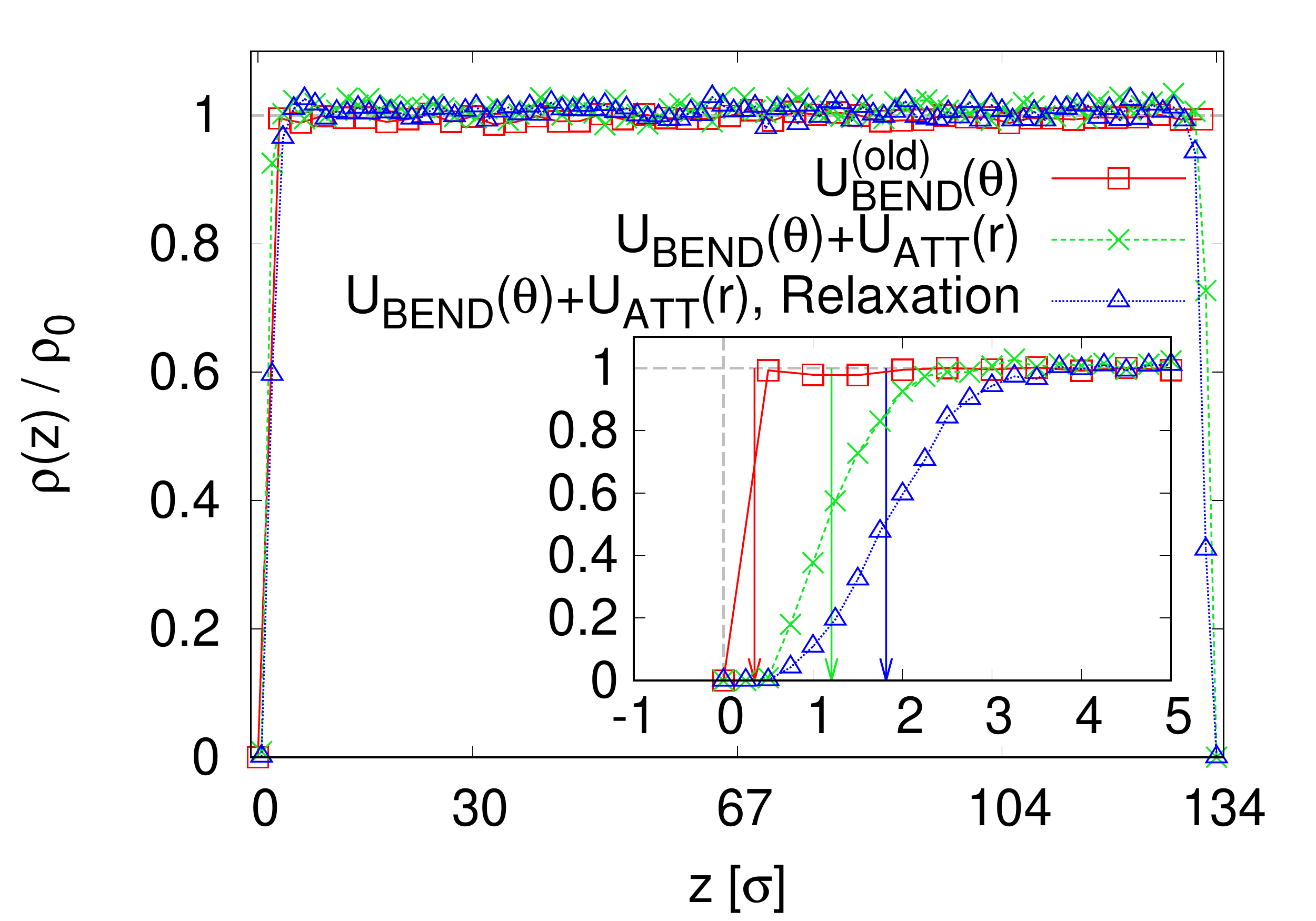}\\
\caption{(a) Time series of diagonal terms of pressure tensor $P_{\alpha \beta}(t)$. (b)(c) Rescaled mean square internal distance, $\langle R^2(s) \rangle / (sl_b^2)$, plotted versus $s$ for all chains (b) and chains having their CMs in the interval $[0.3L_z, 0.7L_z]$. (c) in a confined polymer melt. (d) Monomer density profile rescaled by the bulk melt density $\rho(z)/\rho_0$, plotted as a function of $z$ between two walls located at $z=0\sigma$ and $z=L_z=134\sigma$. Data for confined polymer melts based on two variants of bead-spring model are shown, as indicated. In (b)(c), data for the reference systems (bulk melts), and for the free-standing film after relaxing for $30000\tau$ are also shown for comparison. In the inset of (d), only data for $z\le 5$ are shown. The Gibbs dividing surfaces for the confined film at different status at $z=z_G^{\rm (lower)}=0.35\sigma$, $1.21\sigma$, and $1.82\sigma$ from left to right are marked by arrows. The root mean square bond length increases slightly from $\ell_b \approx 0.964\sigma$ to $\ell_b \approx 0.967\sigma$ after switching the potential $U_{\rm BEND}^{\rm (old)}(\theta)$ to $U_{\rm BEND}(\theta)$ and $U_{\rm ATT}(r)$.}
\label{fig-p0}
\end{center}
\end{figure*}

In order to study free-standing or supported films {at pressure $P=0.0\epsilon/\sigma^{3}$}, stabilizing {non-bonded} attractive monomer monomer interactions are required. For this {we} switch to our recently developed CG model~\cite{Hsu2019,Hsu2019e} based on the bead-spring model~\cite{Kremer1990,Kremer1992} by simply turning on the attractive potential $U_{\rm ATT}(r)$ \{Eq.~(\ref{eq-Uatt})\} {for non-bonded monomer pairs} and replacing the bending potential $U_{\rm BEND}^{\rm (old)}(\theta)$ with  $U_{\rm BEND}(\theta)$ \{Eq.~(\ref{eq-UBEND})\}. This choice of interaction has the additional advantage that it allows us to study glassy films as well. Starting from a fully equilibrated polymer melt confined between two walls obtained in the last section, we perform MD simulations in the NVT ensemble with a Langevin thermostat at the temperature $T=1.0\epsilon/k_B$ using the package ESPResSo++~\cite{Espressopp,Espressopp20}, keeping the short range repulsion from the walls. Fig.~\ref{fig-p0}a shows that the three diagonal terms of the pressure tensor $P_{\alpha \beta}(t)$ drop from $4.9\epsilon/\sigma^3$ ($5.0\epsilon/\sigma^3$ for bulk melts) to $0.1\epsilon/\sigma^3$ in a very short time about $20 \tau$. We further relax the confined polymer film for $30000\tau \approx 13\tau_e$, $\tau_e\approx 2266\tau$ being the entanglement time~\cite{Hsu2016}, by performing MD simulations in the NPT ensemble (Hoover barostat with Langevin thermostat~\cite{Martyna1994,Quigley2004} implemented in ESPResSo++~\cite{Espressopp,Espressopp20}) at temperature $T=1.0\epsilon/k_B$ and pressure $P=(P_{xx}+P_{yy}+P_{zz})/3 =0.0\epsilon/\sigma^3$ to finally adjust the pressure from $0.1\epsilon/\sigma^3$ to $0.0\epsilon/\sigma^3$. Under this circumstance, an equilibrated free-standing film is generated after removing two walls by turning off the wall potential at $z=0\sigma$ and $z=L_z=134\sigma$. If we only remove one of the walls, we get a polymer film with one supporting substrate, where {one} of course can introduce appropriate adhesion interactions.

\begin{figure*}[t!]
\begin{center}
\includegraphics[width=0.85\textwidth,angle=0]{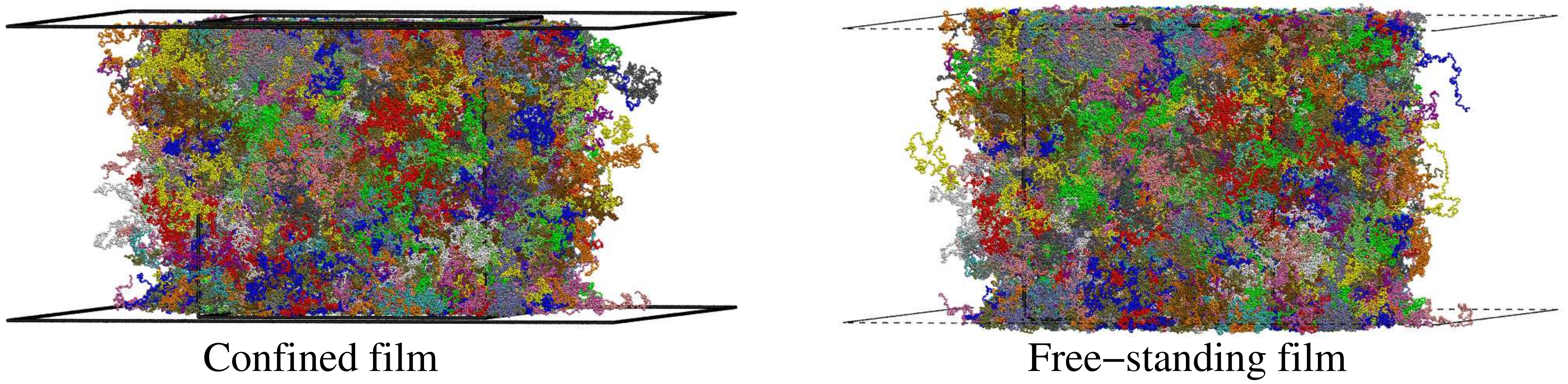}
\caption{Snapshots of the configurations of a fully equilibrated melt containing $n_c=1000$ chains of $N=2000$ monomers confined between two walls (a confined film) at the melt density $\rho=0.85\sigma^{-3}$ and with free surfaces (a free-standing film) at $T=1.0\epsilon/k_B$ and $P=0.0\epsilon/\sigma^3$.}
\label{fig-films}
\end{center}
\end{figure*}

The overall conformations of all chains and inner chains ($0.3L_z \le r^{(i)}_{\rm CM,z} \le 0.7L_z$) as characterized by $\langle R^2 (s) \rangle$ between two walls for a confined polymer melt based on two variants of bead-spring model, and after relaxing for $30000\tau$ (NPT MD simulations) are preserved within fluctuation as shown in Fig.~\ref{fig-p0}b,c. The monomer density profile $\rho(z)$ in the direction perpendicular to the wall compared to the density in the interior of confined melt is also preserved as shown in Fig.~\ref{fig-p0}d. The thickness of films, $D$, is determined according to the concept of Gibbs dividing surface that has been applied to identify the interface between two different phases~\cite{Hansen2013,Kumar1994,Peter2006}, e.g. liquid and vapor, polymer and vacuum, etc. based on the density profile in the direction perpendicular to the interfaces. The locations of the Gibbs dividing surfaces (planar surfaces) corresponding to the upper and lower bounds of films, 
\begin{equation}
     z_G^{\rm (upper)} = z_c + \frac{1}{\bar{\rho}}\int_{z_c}^{z_{\rm max}}\rho(z) dz
\quad {\rm and} \quad 
     z_G^{\rm (lower)} = z_c - \frac{1}{\bar{\rho}}\int_{z_{\rm min}}^{z_c} \rho(z)dz \,
\label{eq-zG}
\end{equation}
are obtained by the requirement of equal areas that 
\begin{equation}
    \int_{z_c}^{z_G^{\rm (upper)}}(\rho(z)-\bar{\rho}) dz = \int_{z_G^{\rm (upper)}}^{z_{\rm max}} (\rho(z)-0) dz
\end{equation}
and
\begin{equation}
    \int_{z_{\rm min}}^{z_G^{\rm (lower)}}(\rho(z)-0) dz = \int_{z_G^{\rm (lower)}}^{z_c} (\rho(z)-\bar{\rho}) dz \,,
\end{equation}
respectively. Here $z_{\rm min}$ and $z_{\rm max}$ are the two limits where $\rho(z)$ approaches to zero, and $z_c=(z_{\rm min}+z_{\rm max})/2$. The mean monomer density is given by
\begin{equation}
        \bar{\rho}=\frac{1}{2\Delta z}\int_{z_c-\Delta z}^{z_c+\Delta z} \rho(z) dz 
\end{equation}
where the value of $\Delta z$ is chosen such that the monomer density profile $\rho(z)$ in the  interval $[z_c-\Delta z,z_c+\Delta z]$ reaches a plateau value within small fluctuation. {Choosing $\Delta = 2.5\sigma$,} the thickness of confined film, $D=z_G^{\rm (upper)}-z_G^{\rm (lower)}$, at $T=1.0\epsilon/k_B$ reduces from $133.4\sigma$ at $P=4.9\epsilon/\sigma^3$ to $131.6\sigma$ at $P=0.1\epsilon/\sigma^3$ due to the short-range attractive interaction between non-bonded monomers, and finally is stabilized at $130.3\sigma$ at $P=0.0\epsilon/\sigma^3$ {where the lateral dimensions of film increase slightly from $L_x=L_y \approx  133.0\sigma$ to $L_x=L_y \approx 134.0\sigma$.}

\begin{figure*}[t!]
\begin{center}
\includegraphics[width=0.32\textwidth,angle=270]{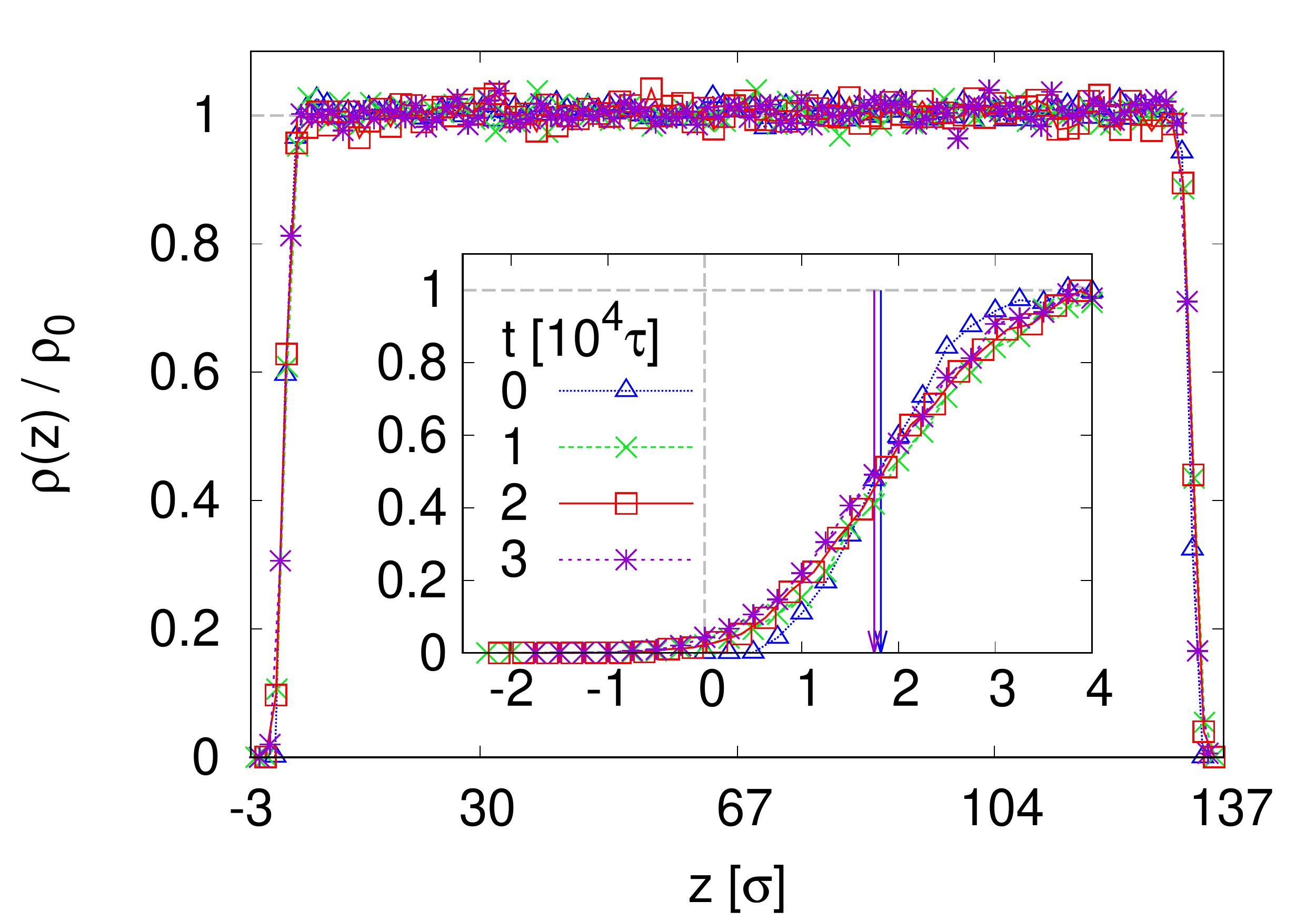}
\caption{Monomer density profile rescaled to the bulk melt density $\rho(z)/\rho_0$, plotted as a function of $z$ along the perpendicular direction of the interfaces of free-standing film at different relaxation times. For comparison, the centers of free-standing and confined films in the $z$-direction are matched. 
In the inset, only data for $z \le 4\sigma$ are shown. The Gibbs dividing surfaces for the confined film at different states at $z=z_G^{\rm (lower)}=1.82\sigma$, $1.75\sigma$ for the relaxation time $t=0\tau$ (right after removing the two walls), and $t=30000\tau$ are marked by arrows.}
\label{fig-free}
\end{center}
\end{figure*}

To further relax the free-standing film, we have also performed MD simulations in the NVT ensemble at the temperature $T=1.0\epsilon/k_B$ for $30000\tau$, where the resulting pressure $P=0.0\epsilon/\sigma^3$. {The perpendicular simulation box size is set to $L_z=194\sigma$ for preventing any interaction between monomers and the lateral surfaces of the box.} Snapshots of the configurations of confined and free-standing films after relaxing for $30000\tau$  are shown in Fig.~\ref{fig-films}. The estimate of $\langle R^2(s) \rangle$ for all chains and chains in the middle part of the free-standing film, and $\rho(z)$ with bin size $0.25\sigma$ are shown in Figs.~\ref{fig-p0}b,c and ~\ref{fig-free}, respectively. We see that after relaxing polymer chains in a free-standing film for $30000\tau \approx 13\tau_e$, the thickness of the free-standing film, $130.5\sigma$, is still compatible with that of the confined polymer film, $130.3\sigma$. The average conformations of all chains and inner chains also remain the same while the tails of the monomer density profile become longer indicating that the surface becomes a bit rougher.

\begin{figure*}[t!]
\begin{center}
(a)\includegraphics[width=0.32\textwidth,angle=270]{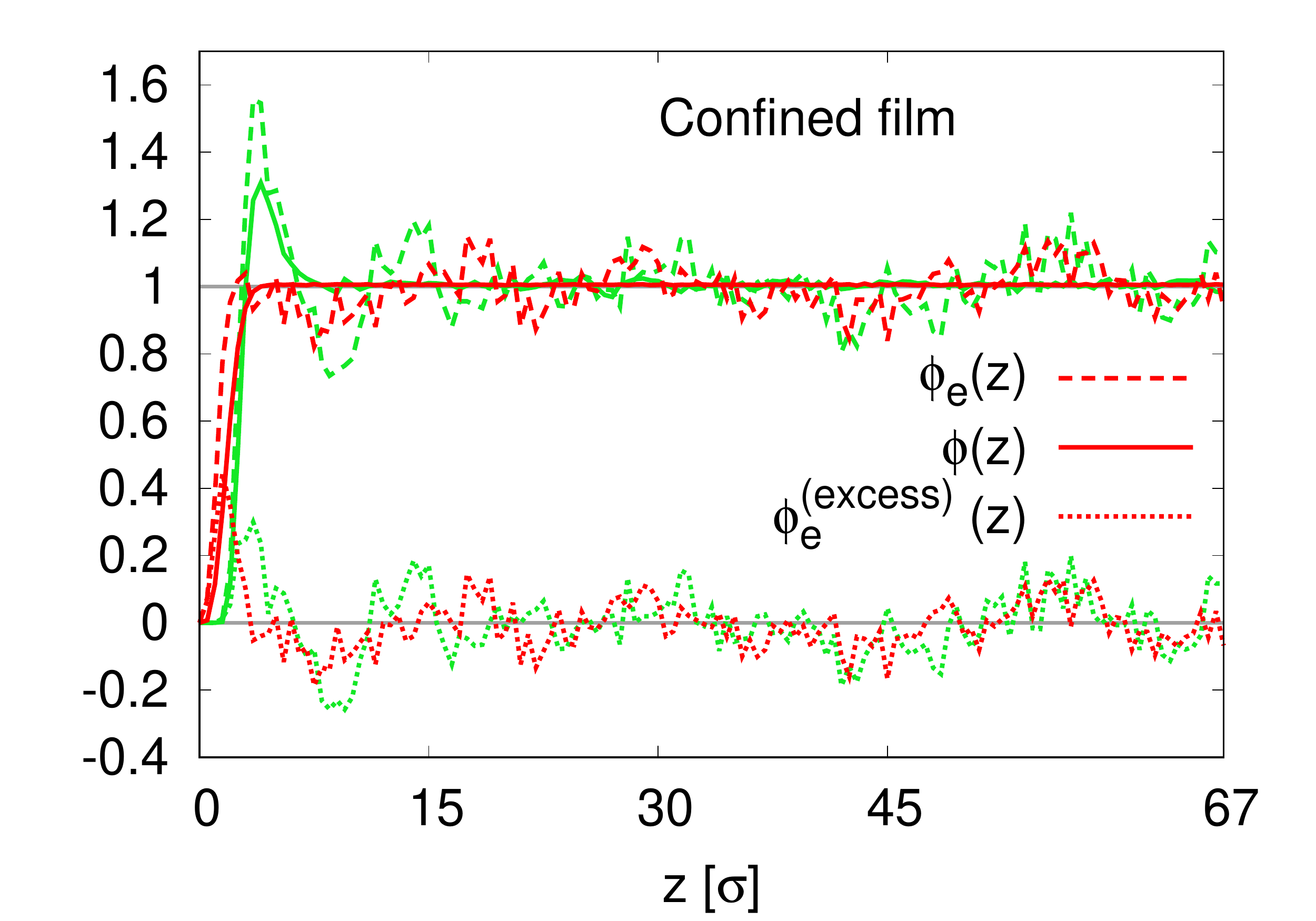}\hspace{0.1cm}
(b)\includegraphics[width=0.32\textwidth,angle=270]{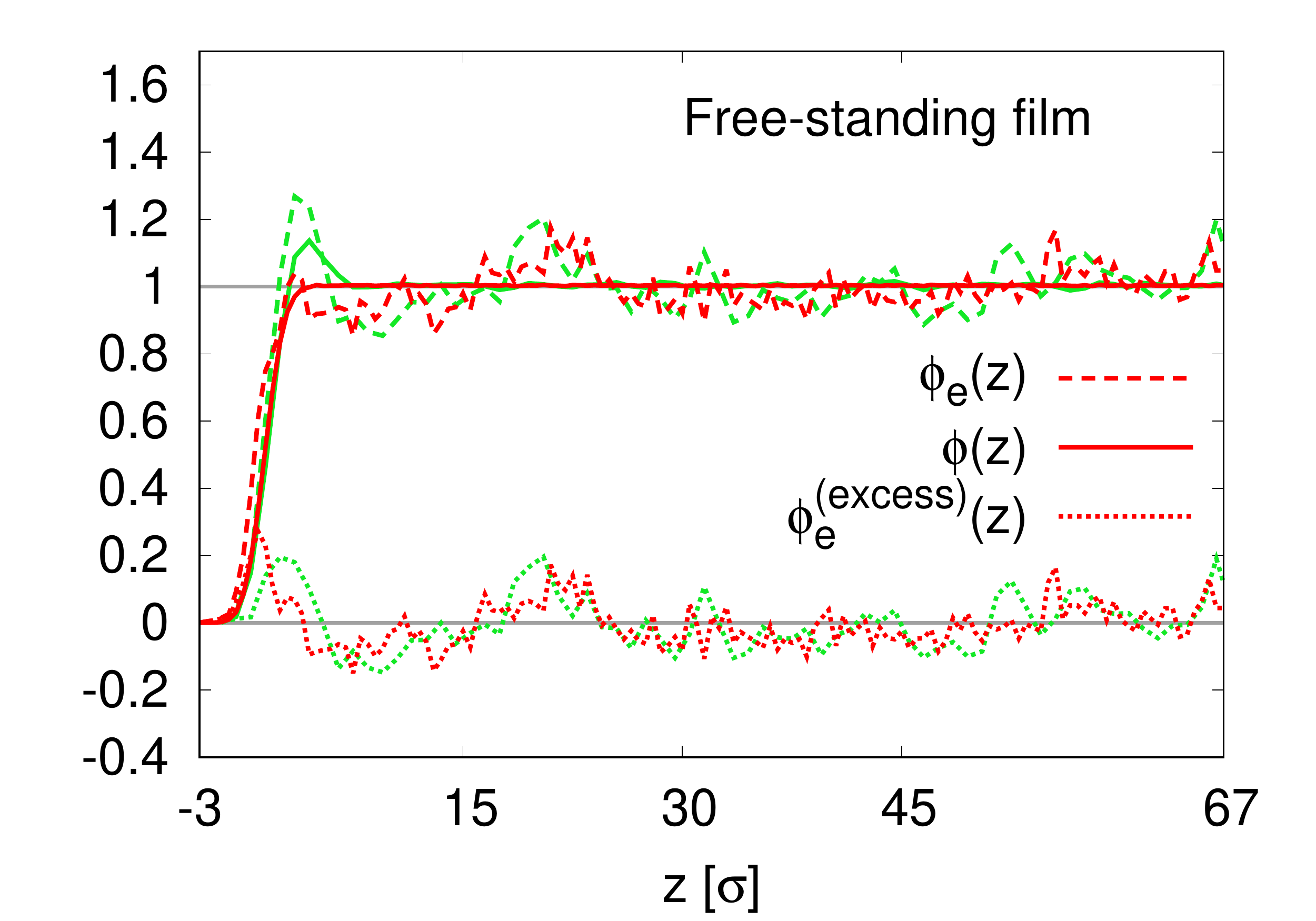}\\
(c)\includegraphics[width=0.43\textwidth,angle=270]{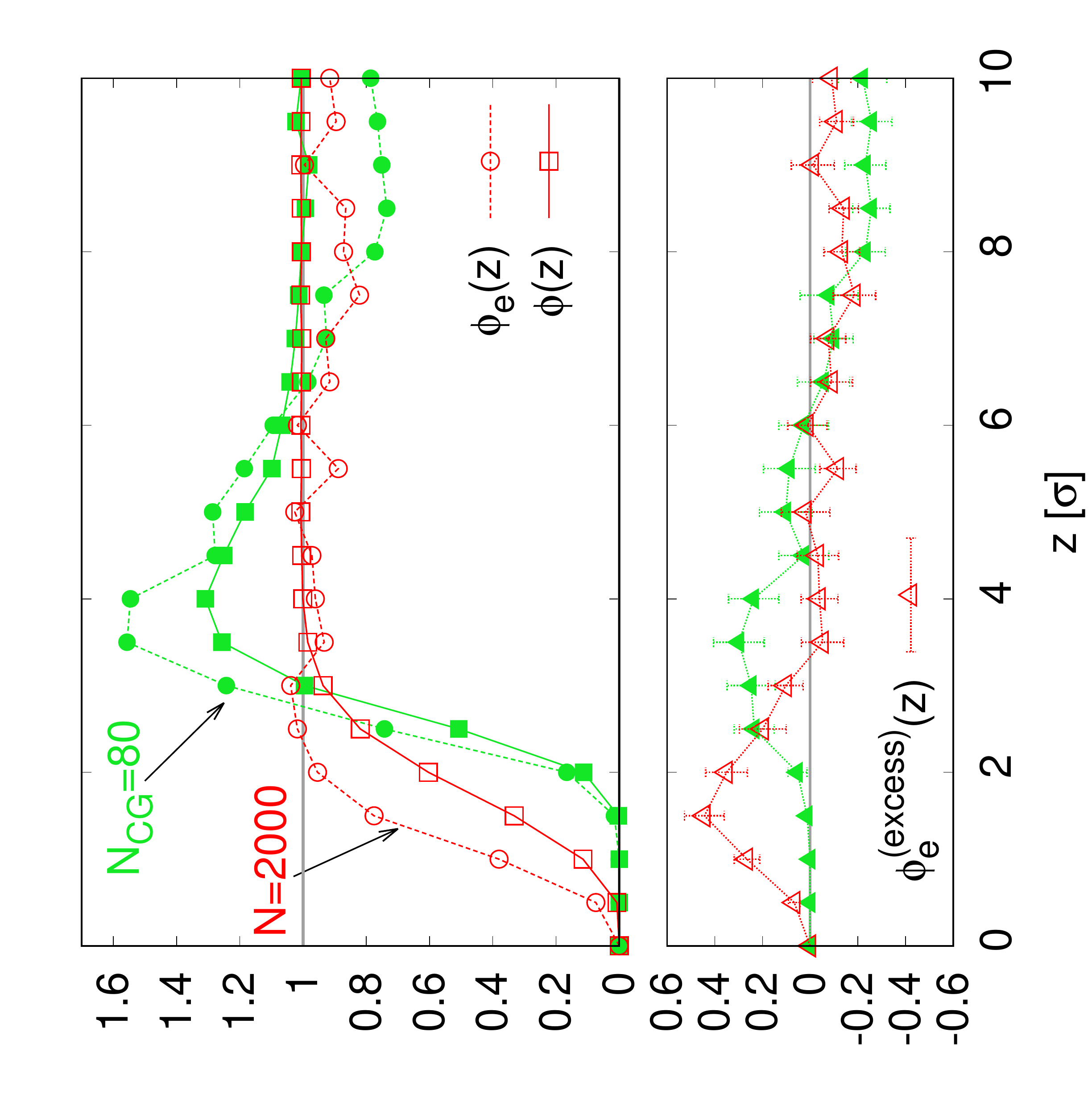}\hspace{0.1cm}
(d)\includegraphics[width=0.43\textwidth,angle=270]{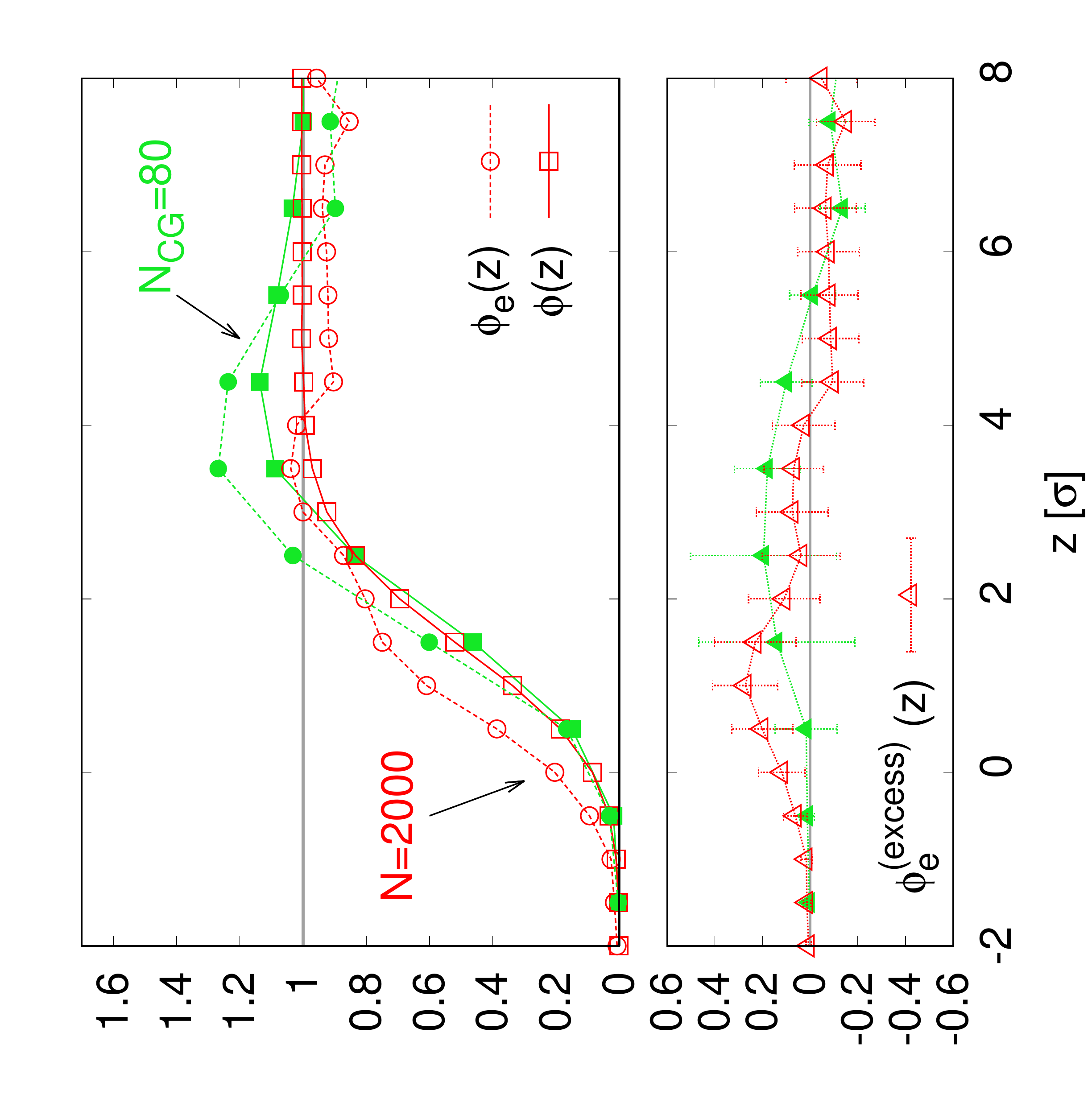}\\
\caption{Normalized densities of all monomers and end monomers, $\phi(z)$ and $\phi_e(z)$, respectively, and relative excess density of end monomers, $\phi_e^{\rm (excess)}(z)$, plotted as a function of $z$ along the perpendicular direction to the interfaces of confined (a)(c) and free-standing films (b)(d). Explicit bead-spring chains are shown by red while the underlying soft-sphere chains are shown in green as indicated. For the most sensitive data $\phi_e^{\textrm{(excess)}}$ error bars are included in (c) and (d).
The centers of free-standing and confined films in the $z$-direction are located at $z=67\sigma$
for comparison.}
\label{fig-pend}
\end{center}
\end{figure*}

Finally, the density of chain ends near surfaces for both confined and free-standing films at $T=1.0\epsilon/k_B$ and $P=0.0\epsilon/\sigma^3$ is examined. For this we compare the normalized density of all monomers, $\phi(z)=\rho(z)/\rho_0$, the density of end monomers, $\phi_e(z)=\rho_e(z)/(2\rho_0/N)$, and the relative excess of end monomers, $\phi_e^{\textrm{(excess)}}(z)=[\rho_e(z)-(2/N)\rho(z)]/(2\rho_0/N)$, along the perpendicular direction of interfaces, averaged over $30$ configurations within $6\tau_e$, are shown in Fig.~\ref{fig-pend}. To illustrate the dependency on chain length or bead size, we have also mapped the bead-spring chains of $N=2000$ monomers of size $\sigma=1$ onto underlying soft-sphere chains of $N_{\rm CG}=80$ spheres of approximate size $6\sigma$ (average radius $\langle R_g(N_b=25) \rangle \approx 3.0\sigma$). 
For films we observe a weak enrichment of end monomers at the surfaces due to a potential gain of entropy~\cite{Wu1995,Matsen2014}. The related depletion zone for chains ends near the surface, as predicted by self-consistent field theories~\cite{Wu1995,Matsen2014},  however turns out to be too small be resolved within the fluctuations of our data. 
The coarse-graining slightly smears out this enrichment effect due to the overlap of soft spheres. For the free-standing film, the enrichment effect of end monomers near the interfaces is slightly less pronounced while at the same time, the interface widens. This indicates a weak roughening of the free surface compared to the confined film. Nevertheless, the indicated $\phi_e^{\rm (excess)}(z)$ near the surfaces in both cases is very small and levels off on the scale of the typical bulk density correlation length given e.g. in Fig.~\ref{fig-eqbsm-wall}.  

\section{conclusion}

In this paper, we have developed an efficient methodology to equilibrate long chain polymer films and applied this method to a polymer film where $1000$ chains of $2000\approx 72 N_e$ monomers are confined between two repulsive walls at the bulk melt density $\rho=0.85\sigma^{-3}$. Starting from a confined CG melt of $1000$ chains of $80$ soft spheres~\cite{Vettorel2010,Zhang2013} at rather high resolution such that each sphere corresponds to $25$ monomers, it takes only $12.8$ hours CPU time on a single processor in Intel 3.6GHz PC to prepare an initial configuration based on the bead-spring model ($10$ hours for equilibrating the confined CG melt using a MC simulation and $2.8$ hours for reinserting monomers into soft spheres using a MD simulation). By gradually switching on the excluded volume interactions between two monomers, overlapping monomers are pushed away slowly in a warm-up procedure. Finally, the confined polymer melt is relaxed with full standard potentials. This takes about $182$ hours CPU time using 48 cores (2.7GHz) on Dual Intel Xeon Platinum 8168 ($155$ hours for the warm-up procedure, $27$ hours for the relaxation procedure). Similarly, as found in the previous studies~\cite{Zhang2014,Ohkuma2018}, the required MD time  for equilibrating confined polymer melts based on the bead-spring model is only about $t=12900\tau \approx 5.69 \tau_e$. 

Following the same strategy, one can easily equilibrate highly entangled polymer melts confined between two walls at distances ranging from thick films to  thin films (in which the distance between two walls is smaller than the radius of gyration of chains) within easily manageable computing time. Our work opens ample possibilities  to study static and dynamic properties of  highly entangled polymer chains in large polymer films, including e.g. entanglement distributions. Varying the interaction potential between walls and monomers, or even replacing the wall potential by other potentials, only requires local short relaxation runs starting from a fully equilibrated polymer melt confined between two repulsive walls. Switching to our recently developed coarse-grained model for studying polymer melts  under cooling~\cite{Hsu2019,Hsu2019e}, both fully equilibrated confined and free-standing films at the temperature $T=1.0\epsilon$ and pressure $P=0.0\epsilon/\sigma^3$ are also obtained in this work. This provides a direct route to further investigate the relation between the glass transition temperature and the thickness of films of highly entangled polymer chains at the zero pressure. Beyond that, it is also interesting to analyze the rheological properties and local morphology of deformed films by stretching or shearing.\\
\\

\noindent
{\bf DATA AVAILABILITY}

The data that support the findings of this study are available from the corresponding author upon 
reasonable request.

\begin{acknowledgments}
H.-P. H. thanks T. Ohkuma and T. Stuehn for helpful discussions.
We also thank K. Ch. Daoulas for carefully reading our paper.
This work has been supported by European Research Council under the European
Union's Seventh Framework Programme (FP7/2007-2013)/ERC Grant Agreement
No.~340906-MOLPROCOMP.
We also gratefully acknowledge the computing time granted by the John von
Neumann Institute for Computing (NIC) and provided on the supercomputer JUROPA
at J\"ulich Supercomputing Centre (JSC).
\end{acknowledgments}
\bibliography{hpkk_eqfilm18092020.bbl}
\end{document}